\documentclass{aa}  

\usepackage{graphicx}
\usepackage{txfonts}
\usepackage[labelformat=simple,justification=centering]{subcaption}

\usepackage[version=4]{mhchem}
\def\HCO+{HCO$^+$}
\def\HxiiiCO+{H$^{13}$CO$^+$}
\def\DCO+{DCO$^+$}

\def\NiiH+{N$_2$H$^+$}
\def\Hii{H$_2$}
\def\Hii+{H$_2^+$}
\def\Hiii+{H$_3^+$}
\def\HiiiO+{H$_3$O$^+$}
\def\e-{e$^-$}

\usepackage{xspace}

\usepackage{tikz}
\usetikzlibrary{shapes,arrows}
\usetikzlibrary{positioning}

\usepackage{tikz}
\def\checkmark{\tikz\fill[scale=0.4](0,.35) -- (.25,0) -- (1,.7) -- (.25,.15) -- cycle;}

\usepackage{cleveref}

\usepackage{tabls}

\usepackage{xcolor}

\usepackage[flushleft]{threeparttable}
\usepackage{tabularx}

\usepackage{gensymb}

\begin{document}

   \title{Chemically tracing the water snowline in protoplanetary disks with HCO$^+$}


   \author{M. Leemker
          \inst{1}
          \and
          M. L. R. van 't Hoff\inst{1,2}
          \and
          L. Trapman \inst{1}
          \and
          M. L. van Gelder\inst{1}
          \and
          M. R. Hogerheijde\inst{1,3}
          \and
          D. Ru{\'\i}z-Rodr{\'\i}guez\inst{4}
          \and
          E. F. van Dishoeck\inst{1,5}}

   \institute{Leiden Observatory, Leiden University, P.O. box 9513, 2300 RA Leiden, The Netherlands\\
              \email{leemker@strw.leidenuniv.nl}
         \and
			 University of Michigan, Department of Astronomy, 1085 S. University, Ann Arbor, MI 48109, USA
    	 \and
    	     Anton Pannekoek Institute for Astronomy, University of Amsterdam, Science Park 904, 1090 GE Amsterdam, The Netherlands
         \and
         	National Radio Astronomy Observatory, 520 Edgemont Road, Charlottesville, VA 22903-2475, USA
         \and
         	 Max-Planck-Institut f\"ur Extraterrestrische Physik, Giessenbachstrasse 1, 85748 Garching, Germany 
             }

 
  \abstract{The formation of planets is expected to be enhanced around snowlines in protoplanetary disks, in particular around the water snowline. Moreover, freeze-out of abundant volatile species in disks alters the chemical composition of the planet-forming material. However, the close proximity of the water snowline to the host star combined with the difficulty of observing water from Earth makes a direct detection of the water snowline in protoplanetary disks challenging. HCO$^+$ is a promising alternative tracer of the water snowline. The destruction of HCO$^+$ is dominated by gas-phase water, leading to an enhancement in the HCO$^+$ abundance once water is frozen out.}
  {Following earlier observed correlations between water and H$^{13}$CO$^+$ emission in a protostellar envelope, the aim of this research is to investigate the validity of HCO$^+$ and the optically thin isotopologue, H$^{13}$CO$^+$, as tracers of the water snowline in protoplanetary disks and the required sensitivity and resolution to observationally confirm this. }
  {A typical Herbig Ae disk structure is assumed and its temperature structure is modelled with the thermochemical code \texttt{DALI}. Two small chemical networks are then used and compared to predict the HCO$^+$ abundance in the disk; one without water and one including water. Subsequently, the corresponding emission profiles are modelled for the $J=2-1$ transition of H$^{13}$CO$^+$ and HCO$^+$, which provides the best balance between brightness and optical depth effects of the continuum emission, and is less affected by blending with complex molecules. Models are then compared with archival ALMA data.}
  {The HCO$^+$ abundance jumps by two orders of magnitude over a radial range of 2~AU outside the water snowline, which in our model is located at 4.5~AU. We find that the emission of H$^{13}$CO$^+$ and HCO$^+$ is ring-shaped due to three effects: destruction of HCO$^+$ by gas-phase water, continuum optical depth, and molecular excitation effects.
  Comparing the radial emission profiles for $J=2-1$ convolved with a $0\farcs 05$ beam reveals that the presence of gas-phase water causes an additional drop of only $\sim$13\%  and 24\% in the center of the disk, for H$^{13}$CO$^+$ and HCO$^+$, respectively. 
    For the much more luminous outbursting source V883 Ori, our models predict that the effects of dust and molecular excitation are not limiting HCO$^+$ as a snowline tracer if the snowline is located at radii larger than $\sim$40~AU. Our analysis of recent archival ALMA band~6 observations of the $J=3-2$ transition of HCO$^+$ is consistent with the water snowline located around 100~AU, further out than was previously estimated from an intensity break in the continuum emission.}
  {The HCO$^+$ abundance drops steeply around the water snowline, when water desorbs in the inner disk, but continuum optical depth and molecular excitation effects conceal the drop in HCO$^+$ emission due to the water snowline. Therefore locating the water snowline with HCO$^+$ observations in disks around Herbig Ae stars is very difficult, but it is possible for disks around outbursting stars such as V883 Ori, where the snowline has moved outwards. 
  }


   \maketitle
%

\section{Introduction}

High resolution observations of protoplanetary disks show that many of them have rings, gaps, and other substructures (e.g., \citealt{Andrews2010, vanderMarel2015, Fedele2017, Huang2018DSHARPII, Long2018}; and \citealt{Andrews2020} for a review). Different explanations have been proposed for these substructures such as planets (e.g. \citealt{Bryden1999, Zhu2014, Dong2018}) and snowlines (e.g. \citealt{Banzatti2015, Zhang2015, Okuzumi2016}). 
A snowline is the midplane radius in a protoplanetary disk where 50\% of one of the major volatiles is in the gas phase and 50\% is frozen out onto the dust grains. These snowlines are related to planet formation, because dust properties change around the H$_2$O and CO or CO$_2$ snowlines \citep{Pinilla2017}. Even though water ice mantles may not always aid dust coagulation by collisions \citep{Kimura2020}, the sublimation, condensation, and diffusion of gas-phase water enhances the surface density around the water snowline aiding planet formation and potentially triggering the streaming instability (e.g. \citealt{Stevenson1988, Drazkowska2017, Schoonenberg2017}). 

Snowlines do not only affect the formation of planets but they also affect their chemical composition. The sequential freeze-out of the major volatile species changes the bulk chemical composition, often measured as the C/O ratio, of the planet-forming material with the ice becoming more oxygen-rich than the gas \citep{Oberg2011, Eistrup2016, Eistrup2018}. The water snowline is of particular importance as water is crucial for the origin of life. 

Knowledge of the water snowline location is thus essential for the understanding of planet formation and composition. 
Yet observing water snowlines is challenging because their expected location is only a few AU from the host star for T~Tauri disks \citep{Harsono2015} and $\sim$10~AU for disks around more luminous Herbig Ae stars. Therefore, observations with very high spatial resolution are necessary. On top of that, observing water snowlines directly from the ground by observing gas-phase water in protoplanetary disks is challenging because of the absorption of water in the Earth's atmosphere. The only detections of water in protoplanetary disks probe the inner most hot parts ($\lesssim$5~AU; e.g. \citealt{Carr2008, Salyk2008, Mandell2012}) or the cold outer parts with space-based telescopes, analysing multiple water lines \citep{Hogerheijde2011, Zhang2013, Blevins2016, Salinas2016, Du2017} or have used both ground based and space based telescopes \citep{Banzatti2017, Salyk2019}. 
Targeting less abundant isotopologues of water, e.g. H$_2^{18}$O, reduces the atmospheric absorption, but detecting these molecules at a significant signal-to-noise ratio greatly increases the required observing time. Space-based telescopes circumvent this problem, but so far, lacked the spatial resolution to resolve the water snowline. 

Chemical tracers provide an alternative way of locating snowlines. Some molecules show ring-shaped emission even if the density of the disk is smooth. One example of such a molecule is CN, which is enhanced if a strong UV field is present (e.g. \citealt{vanZadelhoff2003, Teague2016, Cazzoletti2018}) and other examples are molecules that trace snowlines.
Chemical imaging has been used to locate the CO snowline in the disks around TW~Hya and HD~163296 using N$_2$H$^+$ and DCO$^+$ \citep{Mathews2013, Qi2013, Qi2015, Qi2019}. 
Both tracers show ring-shaped emission but detailed chemical modelling is needed to infer the location of the CO snowline from these observations \citep{Aikawa2015, vantHoff2017, Carney2018}.

Water can be traced chemically using the HCO$^+$ ion, which is destroyed by gas-phase water \citep{Phillips1992, Bergin1998}:
\begin{align}
\mathrm{HCO^+} + \mathrm{H_2O} \to \mathrm{CO} + \mathrm{H_3O^+}. \label{eq:HCOpH2O}
\end{align}
Based on this reaction, HCO$^+$ is expected to be abundant when water is frozen out onto the grains and HCO$^+$ is depleted when water is in the gas phase. This anti-correlation between water and HCO$^+$ has been verified observationally by the optically thin isotopologue H$^{13}$CO$^+$ and an isotopologue of water, H$_2^{18}$O, in a protostellar envelope \citep{vantHoff2018water}. Due to the high luminosity of young stellar objects at this early stage, the snowline is located further away from the star than in protoplanetary disks \citep{Visser2012, Jorgensen2013, Vorobyov2013, Visser2015, Cieza2016, Hsieh2019}. Similarly, the higher luminosity of Herbig Ae stars compared to T~Tauri stars increases the temperature in the protoplanetary disks around them, which locates the water snowline further out in disks around the former type. Therefore, we focus on protoplanetary disks around Herbig Ae stars in this work.

The aim of this paper is to investigate HCO$^+$ as a tracer of the water snowline in protoplanetary disks. Compared to protostellar envelopes the 2D temperature and density structure of disks complicates matters. First, the location of the water snowline in protoplanetary disks around Herbig Ae stars is expected around 10~AU, in contrast to $\gtrsim$100~AU in protostellar envelopes. Second, the higher column density in disks increases the optical depth of the continuum and line emission. This complicates locating the water snowline as emission from HCO$^+$ can be absorbed by dust particles, mimicking the effect of depletion due to gas-phase water on the HCO$^+$ emission. 
Finally, the HCO$^+$ emission can also be ring-shaped when the $J$-level population of the low levels, that can be observed with ALMA, decreases towards the center of the star due to the increase in temperature and density in the inner disk. This also mimics the effect of gas-phase water on the HCO$^+$ emission.

The validity of HCO$^+$ as a tracer of the water snowline in protoplanetary disks is investigated by modelling the physical and chemical structure of a disk around a typical Herbig Ae star with a luminosity of 36~L$_{\odot}$ as described in Section~\ref{sec:models}. 
The predicted abundance structure of HCO$^+$ is presented in Section~\ref{sec:resultschem}, and the corresponding emission profiles of HCO$^+$ and H$^{13}$CO$^+$ are described in Section~\ref{sec:results_rad_cut} and \ref{sec:results_line_profile}. Archival ALMA observations of H$^{13}$CO$^+$ in the outbursting source V883 Ori are discussed in Section~\ref{sec:obs}. Finally we conclude in Section~\ref{sec:concl} that it is difficult to use HCO$^+$ as a tracer of the water snowline in disks around Herbig Ae stars, but that it is possible for outbursting sources such as V883 Ori.

\section{Protoplanetary disk model} \label{sec:models}
The HCO$^+$ and H$^{13}$CO$^+$ radial emission profiles are modelled in several steps. First, a density structure for a typical disk around a Herbig Ae star is assumed. Second, the gas temperature is calculated using the thermochemical code \texttt{DALI} \citep{Bruderer2009, Bruderer2012, Bruderer2013}. The gas density and gas temperature are then used as input for two chemical models, one without water and one with water, that predict the HCO$^+$ abundance in the disk. These abundance profiles together with the density and temperature structure of the disk are used to model the HCO$^+$ and H$^{13}$CO$^+$ emission profiles with \texttt{DALI}.

\subsection{Disk structure} \label{sec:dalimodel}

\begin{table}
\centering
\begin{threeparttable}
	\caption{\texttt{DALI} model parameters for the disk around a typical Herbig Ae star.}    
    \begin{tabularx}{\linewidth}{p{0.5\columnwidth}p{0.5\columnwidth}}
    \hline
        Model parameter & Value  \\ \hline
        \\[-0.7em]
        \textit{Physical structure} &  \\
        $R_{\mathrm{subl}}$ & 0.05 AU \\        
        $R_{\mathrm{c}}$ & 50 AU  \\
		$\Sigma_{\mathrm{c}}$ & 5.8 g$\ \mathrm{cm^{-2}}$ \\    
		$M_{\mathrm{disk}}$ & $0.01 \ \mathrm{M_{\odot}}$  \\   
        $\gamma$ & 1  \\
        $h_{\mathrm{c}}$ & 0.1  \\
        $\psi$ & 0.25 \\
        \\[-0.3em]
        \textit{Dust properties} &  \\
        $\chi$ & 0.2 \\
        $f_{\mathrm{ls}}$ & 0.85  \\
        $\Delta_{\mathrm{gas/dust}}$ & 100  \\
        \\[-0.3em]
        \textit{Stellar spectrum}$^{(1)}$ &  \\     
        Type & Herbig \\
		$L_{\star}$ & 36 $\mathrm{L_{\odot}}$ \\
        $L_{\mathrm{X}}$ & 8(28) erg s$^{-1}$  \\
        $T_{\mathrm{eff}}$ & $1(4)$ K\\
        $T_{\mathrm{X}}$ & 7(7) K \\  
        $\zeta_{\mathrm{c.r.}}$ & 5(-17) $\mathrm{s^{-1}}$ \\ 
        \\[-0.3em]
        \textit{Stellar properties} &  \\          
        $M_{\star}$ & $2.3 \ \mathrm{M_{\odot}}$ \\
        \\[-0.3em]
        \textit{Observational geometry} &  \\
        $i$ & 38$\degree$ \\
        $d$ & 100 pc \\ \hline
    \end{tabularx}
    \begin{tablenotes}
      \small
      \item \textbf{Notes.} $a(b)$ represents $a\times 10^b$. $^{(1)}$ Spectrum of HD~100546 \citep{Kama2016}, which is well approximated by a 10$^4$ K black body spectrum. 
    \end{tablenotes}
    \label{tab:paramsdali}
    \end{threeparttable}
\end{table}

The density structure of the disk is modelled with the thermochemical code \texttt{DALI} following the approach of \citet{Andrews2011}. This approach is based on the self-similar solution for a viscously evolving disk, where the gas surface density of the disk outside the dust sublimation radius follows a power law with an exponential taper \citep{LyndenBell1974, Hartmann1998}:
\begin{align}
\Sigma_{\mathrm{gas}} (R) = \Sigma_{\mathrm{c}} \left (\frac{R}{R_{\mathrm{c}}} \right )^{-\gamma} \exp \left [-\left (\frac{R}{R_{\mathrm{c}}}\right )^{2-\gamma} \right ],
\end{align}
with $\Sigma_{\mathrm{c}}$ the gas surface density at the characteristic radius $R_{\mathrm{c}}$ and $\gamma$ the power law index. A sublimation radius of $R_{\mathrm{subl}} = 0.05 $ AU is assumed. Inside this radius, the surface density is set to 7$\times$10$^2$~cm$^{-3}$. 
An overview of all parameters used for the density and temperature structure can be found in Table~\ref{tab:paramsdali}.

 \begin{figure}
   \centering
  \begin{subfigure}{0.99\columnwidth}
  \centering
  \includegraphics[width=1\linewidth]{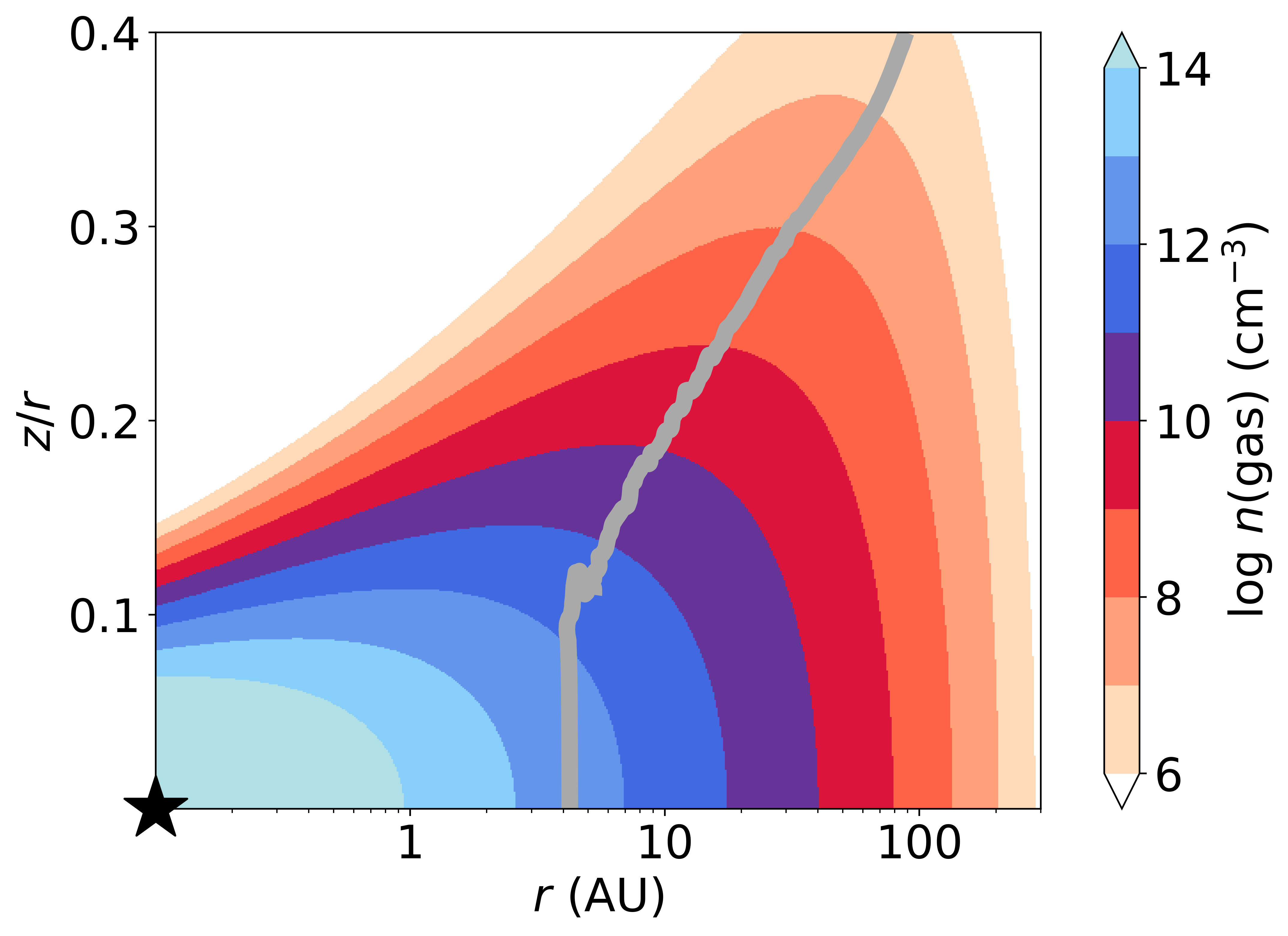}
\end{subfigure}%

\begin{subfigure}{0.99\columnwidth}
  \centering
    \includegraphics[width=1\linewidth]{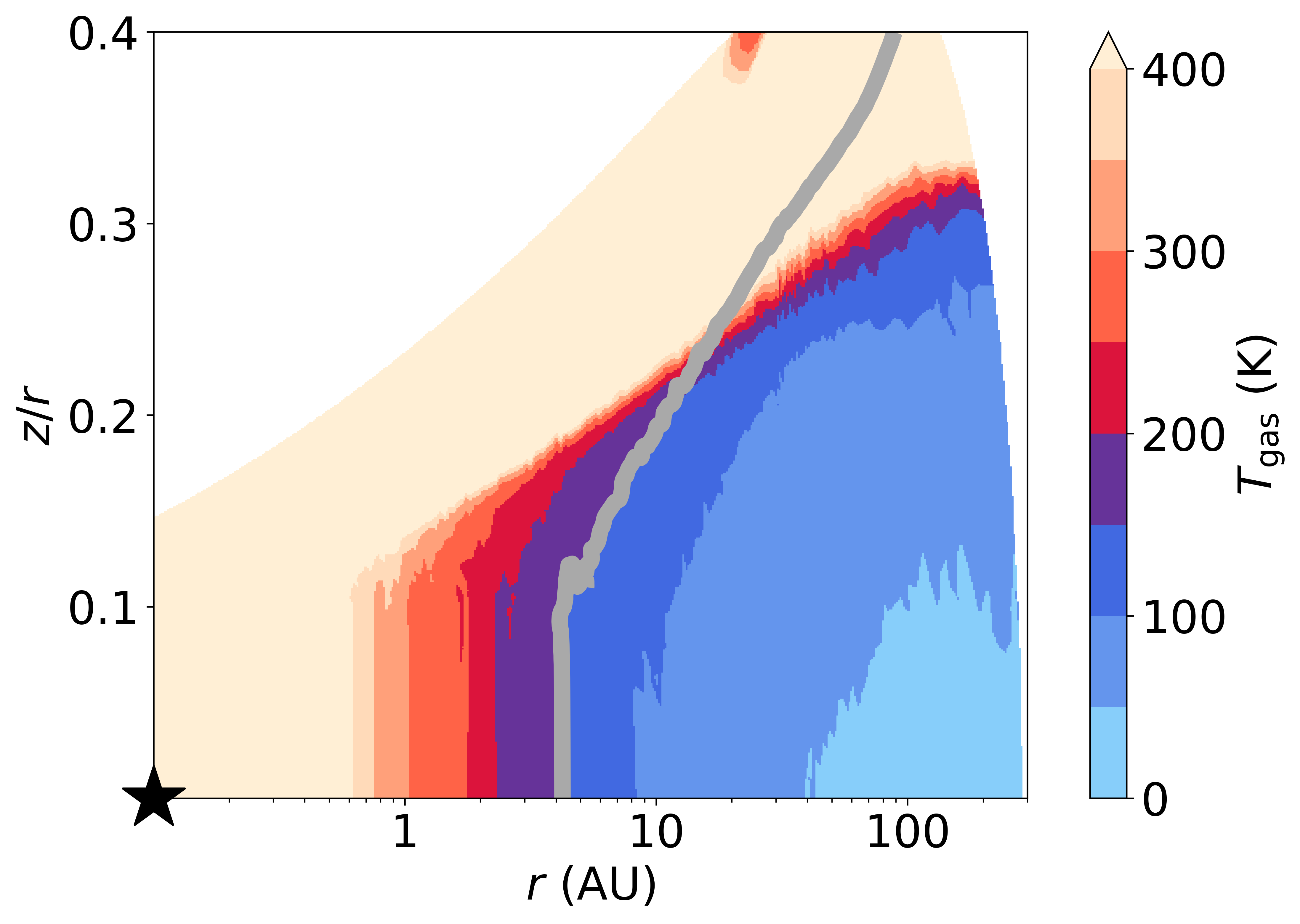}
\end{subfigure}

\begin{subfigure}{0.99\columnwidth}
  \centering
    \includegraphics[width=1\linewidth]{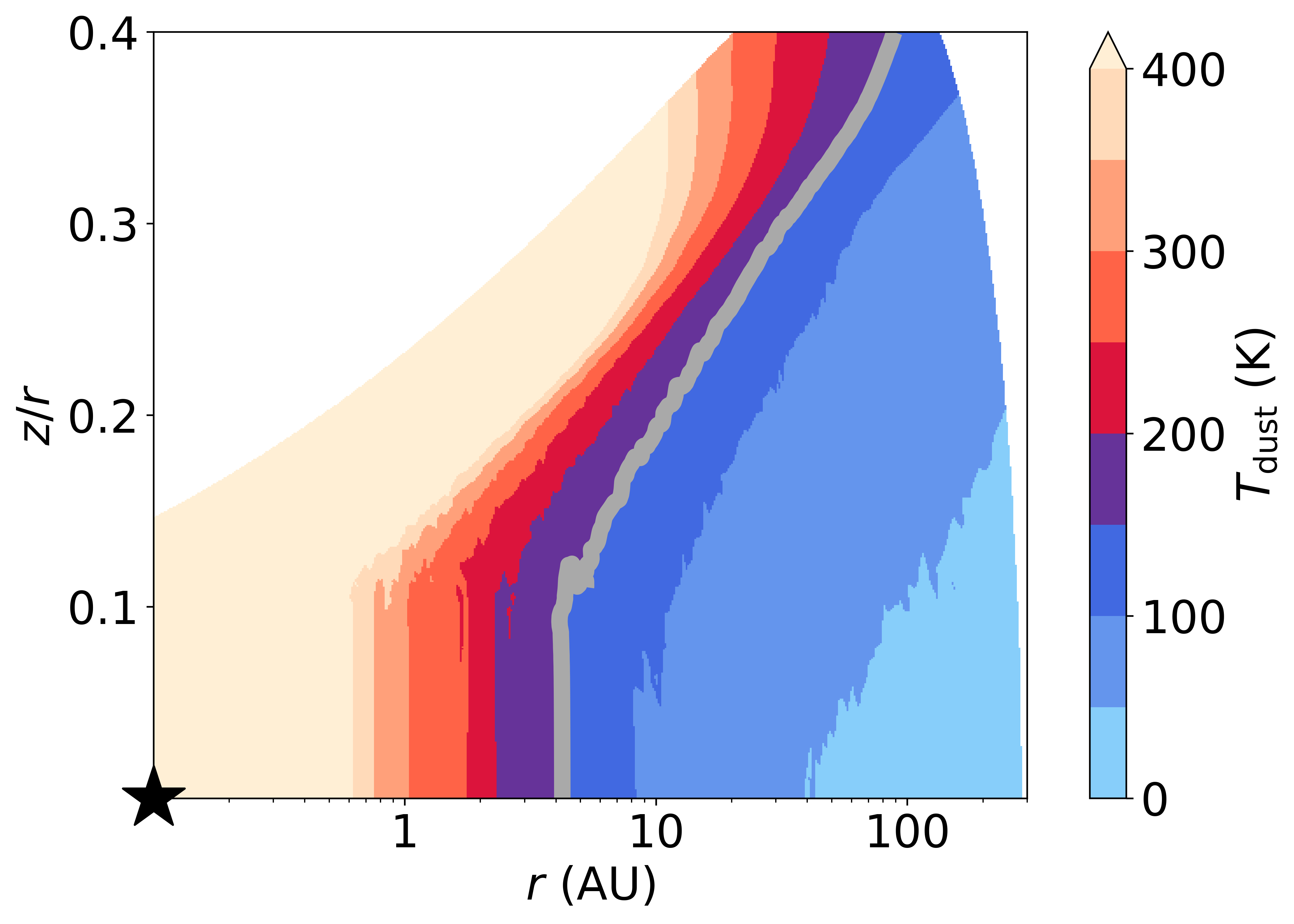}
\end{subfigure}

      \caption{Assumed gas density (top) and computed gas and dust temperature (middle and bottom panels) in the \texttt{DALI} model for a typical disk around a Herbig Ae star with a luminosity of 36~L$_{\odot}$. Only the region where $n(\mathrm{gas}) > 10^6~\mathrm{cm^{-3}}$ is shown. The position of the star is indicated with the symbol of a black star and the $T_{\mathrm{dust}}=150$~K line, approximating the water snow surface, is indicated with the grey solid line. }
         \label{fig:DALImodel}
   \end{figure}

The gas follows a Gaussian distribution in the vertical direction, where the temperature is calculated explicitly at each location in the disk. The scale height of the gas is set by the flaring index $\psi$ and the characteristic scaleheight $h_{\mathrm{c}}$ at $R_{\mathrm{c}}$,
\begin{align}
h = h_{\mathrm{c}} \left ( \frac{R}{R_{\mathrm{c}}} \right )^{\psi}.
\end{align}
The resulting gas density of our model is shown in the top panel of Fig. \ref{fig:DALImodel}. 

The dust surface density is modelled with \texttt{DALI} by scaling the gas surface density with the disk-averaged gas-to-dust mass ratio, $\Delta_{\mathrm{gas/dust}}$, which is set to the ISM value of 100. The vertical structure of the dust is modelled using two populations of dust grains following the approach of \citet{DAlessio2006}. Small grains with sizes between 5~nm and 1~$\mu$m are well mixed with the gas and therefore follow the same vertical structure of the gas. Large grains with sizes from 5~nm to 1~mm are settled to the disk midplane. This is modelled by reducing the scale height of the large grains with a factor $\chi < 1$. The fraction of the dust mass in large grains is controlled by $f_{\mathrm{ls}}$ and the size distribution of both grain populations is assumed to be proportional to $a^{-3.5}$ with $a$ the size of the grains (MRN distribution, \citealt{Mathis1977}). 

The gas temperature needs to be computed separately as the abundances of the molecules that act as coolants of the gas in the disk affect the gas temperature, which in turn affects the chemistry. Therefore, \texttt{DALI} solves for the gas temperature by iterating over the chemistry and heating by the photoelectric effect and the cooling by molecules in the gas. The dust temperature is computed using Monte Carlo continuum radiative transfer. The gas and dust temperature calculated by \texttt{DALI} are shown in the middle and bottom panel of Fig.~\ref{fig:DALImodel}. The gas temperature exceeds the dust temperature in the surface layers only, whereas the gas and dust temperatures are equal in deeper layers of the disk.

\subsection{Chemical models} \label{sec:chemmod} 
\tikzstyle{block} = [rectangle, draw, fill=blue!20, 
    text width=4.2em, text centered, rounded corners, minimum height=2em]
\tikzstyle{newblock} = [rectangle, draw, ultra thick, fill=blue!20, 
    text width=4.2em, text centered, rounded corners, minimum height=2em]

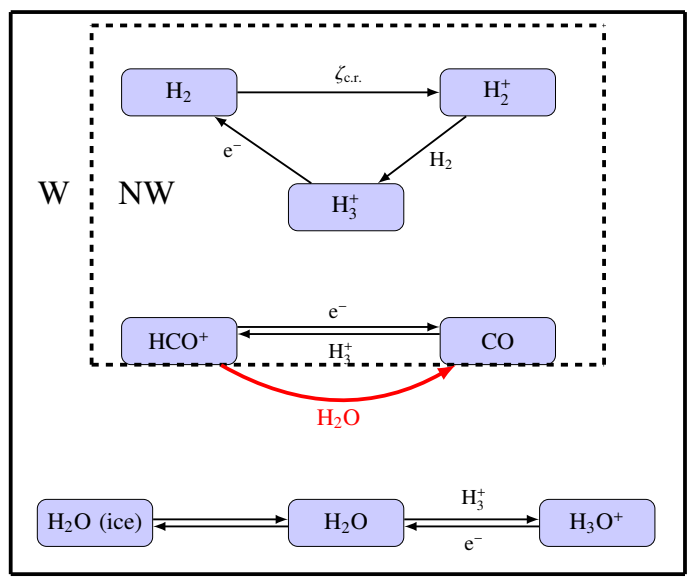
\begin{figure}
\resizebox{9cm}{!}{%
\begin{tikzpicture}[node distance = 1cm, auto] 
    \node [block] (H2) {$\mathrm{H_2}$};
    \node [block, right = 3cm of H2] (H2+) {$\mathrm{H_2^+}$};
    \node [block, below right = 1cm and 0.75cm of H2] (H3+) {$\mathrm{H_3^+}$};
    \node [block, below = 4cm of H3+] (H2O) {$\mathrm{H_2O}$};
    \node [block, left  = 2cm of H2O] (H2Oice) {$\mathrm{H_2O\ (ice)}$};
    \node [block, right = 2cm of H2O] (H3O+) {$\mathrm{H_3O^+}$};
    \node [block, below = 3cm of H2] (HCO+) {$\mathrm{HCO^+}$};
    \node [block, right = 3cm of HCO+] (CO) {CO};
    \draw [dashed, ultra thick] (-1.3,-4.07) -- (-1.3,1);
    \draw [dashed, ultra thick] (6.3,-4.07)  -- (6.3,1);
    \draw [dashed, ultra thick] (-1.3,1)  -- (6.3,1);
    \draw [dashed, ultra thick] (-1.3,-4.07)  -- (6.3,-4.07);    
    \draw [solid, ultra thick] (-2.5,-7.2) -- (-2.5,1.2);
    \draw [solid, ultra thick] (7.5,-7.2)  -- (7.5,1.2);
    \draw [solid, ultra thick] (-2.5,1.2)  -- (7.5,1.2);
    \draw [solid, ultra thick] (-2.5,-7.2)  -- (7.5,-7.2); 
    \node[text width=3cm] at (-0.6,-1.5) {\Large W}; 
    \node[text width=3cm] at (0.6,-1.5) {\Large NW};   
    \draw[-latex, thick] (H2) -- (H2+) node[above,pos=0.5,font=\footnotesize] {$\ \ \ \ $\\[-0.5em]$\zeta_{\mathrm{c.r.}}$};
    \draw[-latex, thick] (H3+) -- (H2) node[left,pos=0.5,font=\footnotesize] {$\mathrm{e^-}\ \ $\\[-0.5em]$ $};
    \draw[-latex, thick] (H2+) -- (H3+) node[below,pos=0.4,font=\footnotesize] {$\ \ \ \ $\\[-0.5em]$\mathrm{H_2}$};
    \draw[-latex, thick] ([yshift=+1.5pt]H2Oice.east) -- ([yshift=+1.5pt]H2O.west) node[midway, below right=5pt,align=left,font=\footnotesize]{};
    \draw[-latex, thick] ([yshift=-1.5pt]H2O.west) -- ([yshift=-1.5pt]H2Oice.east) node[midway, below right=5pt,align=left,font=\footnotesize]{};
    \draw[-latex, thick] ([yshift=-1.5pt]H3O+.west) -- ([yshift=-1.5pt]H2O.east) node[below,pos=0.5,font=\footnotesize] {$\ $\\[-0.5em]$\mathrm{e^-}$};
    \draw[-latex, thick] ([yshift=1.5pt]H2O.east) -- ([yshift=+1.5pt]H3O+.west) node[above,pos=0.5,font=\footnotesize] {$\ $\\[-0.5em]$\mathrm{H_3^+}$};
    \draw[-latex, thick] ([yshift=+6pt]HCO+.east) -- ([yshift=+6pt]CO.west) node[above,pos=0.5,font=\footnotesize]{$\mathrm{e^-}$};
    \draw[-latex, thick] ([yshift=+3pt]CO.west) -- ([yshift=+3pt]HCO+.east) node[below, pos=0.5,font=\footnotesize]{$\ \mathrm{H_3^+}$};
    \path[-latex, ultra thick, red] (HCO+) edge[bend right] node [below] {$\mathrm{H_2O}$} (CO);
    
\end{tikzpicture}
}
\caption{A schematic view of chemical networks NW (no water) and W (water) used to predict the abundance of HCO$^+$. 
The first network, network NW, only includes the reactions enclosed in the dashed box. The second network, network W, includes all reactions present in the figure. The destruction of HCO$^+$ by gas-phase water is indicated with the thick red arrow.}
       \label{fig:networkC}
\end{figure}

\begin{table*}[]
\begin{threeparttable}
    \centering
	\caption{Reaction coefficients of the reactions used in this paper. The last two columns indicate in which chemical networks the reactions are used.}    
     \begin{tabularx}{\linewidth}{p{0.05\columnwidth}p{0.44\columnwidth}p{0.2\columnwidth}p{0.08\columnwidth}p{0.08\columnwidth}p{0.21\columnwidth}p{0.21\columnwidth}p{0.11\columnwidth}p{0.09\columnwidth}p{0.03\columnwidth}p{0.05\columnwidth}}    
    \hline
        & reaction & $\alpha$ & $\beta$ &$\gamma$ & $k (100~\mathrm{K})$ & $k (150~\mathrm{K})$ & units & ref. & NW & W  \\ \hline
       $\zeta_{\mathrm{c.r.}}$ & $\mathrm{H_2 + c.r. \to H_2^+ + e^-}$ & $1(-17)$ & $0$ & $0$ & $1(-17)$ & $1(-17)$& $\mathrm{s^{-1}}$ &  & \checkmark & \checkmark \\         
       $k_{\mathrm{2}}$ & $\mathrm{H_2^+ + H_2 \to H_3^+ + H}$ & $2.1(-9)$ & $0$ & $0$ & $2.1(-9)$ & $2.1(-9)$ & $\mathrm{cm^3\ s^{-1}}$ & (a) & \checkmark & \checkmark \\         
       $k_{\mathrm{3}}$ & $\mathrm{H_3^+ + e^- \to H_2 + H}$ & $2.3(-8)$& $-0.5$ & $0$ & $4.1(-8)$ & $3.4(-8)$ & $\mathrm{cm^3\ s^{-1}}$ & (b) & \checkmark & \checkmark  \\         
       $k_{\mathrm{4}}$ & $\mathrm{CO + H_3^+ \to HCO^+ + H_2}$ & $1.4(-9)$ & $-0.1$ & $-3.4$ & $1.6(-9)$ & $1.5(-9)$ & $\mathrm{cm^3\ s^{-1}}$ & (c) & \checkmark & \checkmark  \\         
       $k_{\mathrm{e^-}}$ & $\mathrm{HCO^+ + e^- \to CO + H}$ & $2.4(-7)$ & $-0.7$ & $0$ &  $5.1(-7)$ & $3.9(-7)$ & $\mathrm{cm^3\ s^{-1}}$ & (d) & \checkmark & \checkmark \\         
       $k_{\mathrm{H_2O}}$ & $\mathrm{HCO^+ + H_2O \to CO + H_3O^+}$ & $2.5(-9)$ & $-0.5$ & $0$ & $4.3(-9)$ & $3.5(-9)$ & $\mathrm{cm^3\ s^{-1}}$ & (e) && \checkmark \\         
       $k_{\mathrm{7}}$ & $\mathrm{H_3O^+ + e^- \to H_2O + H}$ & $7.1(-8)$ & $-0.5$ & $0$ & $1.2(-7)$& $1.0(-7)$ & $\mathrm{cm^3\ s^{-1}}$ & (f) && \checkmark \\ 
       $k_{\mathrm{8}}$ & $\mathrm{H_2O + H_3^+ \to H_3O^+ + H_2}$ & $5.9(-9)$ & $-0.5$ & $0$ & $1.0(-8)$ & $8.3(-9)$ & $\mathrm{cm^3\ s^{-1}}$ & (g, h) && \checkmark \\    
	   $k_{\mathrm{f}}$ & $\mathrm{H_2O  \to H_2O\ (ice)}$ & $ \dfrac{1.2(-5)n(\mathrm{H_2})}{10(12) \mathrm{cm^{-3}}}$ & $0.5$ & $0$ & $ \dfrac{6.7(-6)n(\mathrm{H_2})}{10(12) \mathrm{cm^{-3}}}$& $ \dfrac{8.3(-6)n(\mathrm{H_2})}{10(12) \mathrm{cm^{-3}}}$ & $\mathrm{s^{-1}}$ & (i) & &  \checkmark \\    
       $k_{\mathrm{d}}$ & $\mathrm{H_2O\ (ice) \to H_2O }$ & $2.8(12)$ & $0$ & $5.8(3)$ & $2.4(-13)$ & $5.4(-5)$ & $\mathrm{s^{-1}}$ & (i) & & \checkmark \\ 
       \hline        
    \end{tabularx}
    \begin{tablenotes}
      \small
      \item \textbf{Notes.} $a(b)$ represents $a\times 10^b$. NW is the chemical network without water, whereas W includes water. All reactions are given in the form $k~=~\alpha~(T_{\mathrm{gas}}/300\mathrm{K})^{\beta}\exp(-\gamma/T_{\mathrm{gas}})$, where the values of $\alpha$, $\beta$, and $\gamma$ for $k_{\mathrm{2}}$, $k_{\mathrm{3}}$, $k_{\mathrm{4}}$, $k_{\mathrm{e^-}}$, $k_{\mathrm{H_2O}}$, $k_{\mathrm{7}}$ and $k_{\mathrm{8}}$ are taken from the rate12 UMIST database \citep{McElroy2013} and $k_{\mathrm{d}}$ uses $T_{\mathrm{dust}}$ instead of $T_{\mathrm{gas}}$. References: (a) \citet{Theard1974}, (b) \citet{McCall2004}, (c) \citet{Klippenstein2010}, (d) \citet{Mitchell1990}, (e) \citet{Adams1978}, (f) \citet{Novotny2010}, (g) \citet{Kim1974}, (h) \citet{Anicich1975} and (i) Appendix~\ref{sec:fdwater} and references therein. 
    \end{tablenotes}
    \label{tab:reactions}
\end{threeparttable}
\end{table*}

The HCO$^+$ abundance with respect to the total number of hydrogen atoms in the disk is modelled with a small chemical network, which is solved time-dependently up to 1~Myr with the python function \texttt{odeint}\footnote{The \texttt{odeint} function is part of the Scipy package in python and uses the LSODA routine from the ODEPACK library in FORTRAN.}. This allows us to more easily study the effect of different parameters than with a computationally more expensive full chemical model.
The HCO$^+$ abundance in protoplanetary disks is mainly controlled by three reactions. 
The main formation route of HCO$^+$ involves gas-phase CO:
\begin{align}
\mathrm{CO} + \mathrm{H_3^+} \to \mathrm{HCO^+} + \mathrm{H_2}. \label{eq:HCOpform}
\end{align}
Here, H$_3^+$ is produced by cosmic ray ionization of molecular hydrogen at a rate of $\zeta_{\mathrm{c.r.}}$.
On the other hand, HCO$^+$ is destroyed by gas-phase water (Eq.~\ref{eq:HCOpH2O}) and by dissociative recombination with an electron:
\begin{align}
\mathrm{HCO^+} + \mathrm{e^-} \to \mathrm{CO} + \mathrm{H}. \label{eq:HCOpem}
\end{align}
Therefore, a jump in the HCO$^+$ abundance around the water snowline is expected if electrons are not the dominant destruction mechanism of HCO$^+$. 

To investigate the relationship between HCO$^+$ and H$_2$O, two small chemical networks are used.   
The first network, network \textit{no water} (NW), is shown in the dashed box in Fig.~\ref{fig:networkC}. This network
includes three reactions to model the ionization in the disk (reactions $\zeta_{\mathrm{c.r.}}$, $k_{\mathrm{2}}$ and $k_{\mathrm{3}}$ in Table~\ref{tab:reactions}), and two reactions to model the HCO$^+$ abundance in the absence of gas-phase water (reactions $k_{\mathrm{4}}$ and $k_{\mathrm{e^-}}$ (Eq.~\ref{eq:HCOpform} and \ref{eq:HCOpem}) in Table~\ref{tab:reactions}). As reaction $k_{\mathrm{H_2O}}$ (Eq.~\ref{eq:HCOpH2O}) is not included in this network, HCO$^+$ is not expected to trace the water snowline in network NW. Therefore, this network serves as a baseline to quantify the effect of the gas-phase water on the HCO$^+$ abundance.
The second network, network \textit{water} (W), is shown in the solid box in Fig.~\ref{fig:networkC} and contains all reactions present in Table~\ref{tab:reactions}. Network W includes those in network NW together with reactions to model the effect of water on the HCO$^+$ abundance. The most important reaction for our study is indicated in red in Fig.~\ref{fig:networkC} and is the destruction of HCO$^+$ by gas-phase water (reaction $k_{\mathrm{H_2O}}$ or Eq.~\ref{eq:HCOpH2O}). The other reactions present in network W include the freeze-out and desorption of water (reactions $k_{\mathrm{f}}$ and $k_{\mathrm{d}}$) and the formation and destruction of gas-phase water from and to H$_3$O$^+$ (reactions $k_{\mathrm{7}}$ and $k_{\mathrm{8}}$). Therefore, HCO$^+$ is expected to trace the water snowline in network W. The equations for the freeze-out and desorption rates of water can be found in Appendix~\ref{sec:fdwater}.

Initially, the abundance of all gas- and ice-phase species are set to 0 except for the abundance of gas-phase CO, which is set to $10^{-4}$ both in network NW and W, and the abundance of gas-phase water in network W, which is set to $3.8\times 10^{-7}$, appropriate for a dark cold cloud \citep{McElroy2013}, see also Table~\ref{tab:params_chem_net}. The rate coefficient for the freeze-out of water depends on multiple parameters, including the number density and size of the grains. Therefore, we assume in chemical network NW and W a typical grain number density of 10$^{-12}\times n(\mathrm{H_2})$ and a grain size of 0.1~$\mu$m, which set the surface area available for chemistry on grains. Assuming a single grains size for the chemistry is an approximation, but the effect of grain growth in disks, which decreases this area, is cancelled by the dust settling in disks, which increases this area as there is more dust available for chemistry in the midplane \citep{Eistrup2016}. 
Finally, the H$^{13}$CO$^+$ abundance is taken to be a factor 70 smaller than the abundance of HCO$^+$, corresponding to the typical $^{12}$C/$^{13}$C isotope ratio \citep{Milam2005}.

\subsection{Radiative transfer} \label{sec:raytracing}

Predictions for multiple transitions of HCO$^+$ and H$^{13}$CO$^+$ were made using the ray tracer in \texttt{DALI}, where the outputs of the chemical networks discussed in Section~\ref{sec:chemmod} were used to set the abundance and both line and continuum optical depth are included. The line radiative transfer does not assume LTE. Instead, the excitation is calculated explicitly, where we used the collisional rate coefficients in the LAMDA database \citep{Botschwina1993, Flower1999, Schoier2005}. 

Furthermore, the disk is assumed to be located at a distance of 100~pc and the inclination is taken to be 38$\degree$. Changing the inclination does not significantly change our results. The radial emission profiles are taken to be along the semi-major axis of the disk. The focus in the paper is on the $J=2-1$ transition at 178.375~GHz and 173.507~GHz for HCO$^+$ and H$^{13}$CO$^+$ respectively, because this transition provides the best balance between the brightness of the line and the effects of the continuum optical depth. Moreover, the higher transitions could be blended with emission from complex organic molecules. Predictions for the weaker $J=1-0$ transition, as well as the brighter $J=3-2$ and $J=4-3$ transitions, that are more affected by the optical depth of the continuum emission and emission from complex organic molecules, can be found in Appendix~\ref{app:Radialcutsotherlines}. To mimic high resolution ALMA observations, the emission is convolved with a $0\farcs 05$ beam, unless denoted otherwise.


\section{Modelling results} \label{sec:results}

\subsection{Chemistry} \label{sec:resultschem} 

\begin{figure}
   \begin{subfigure}{1\columnwidth}
   \centering
   \includegraphics[width=1\linewidth]{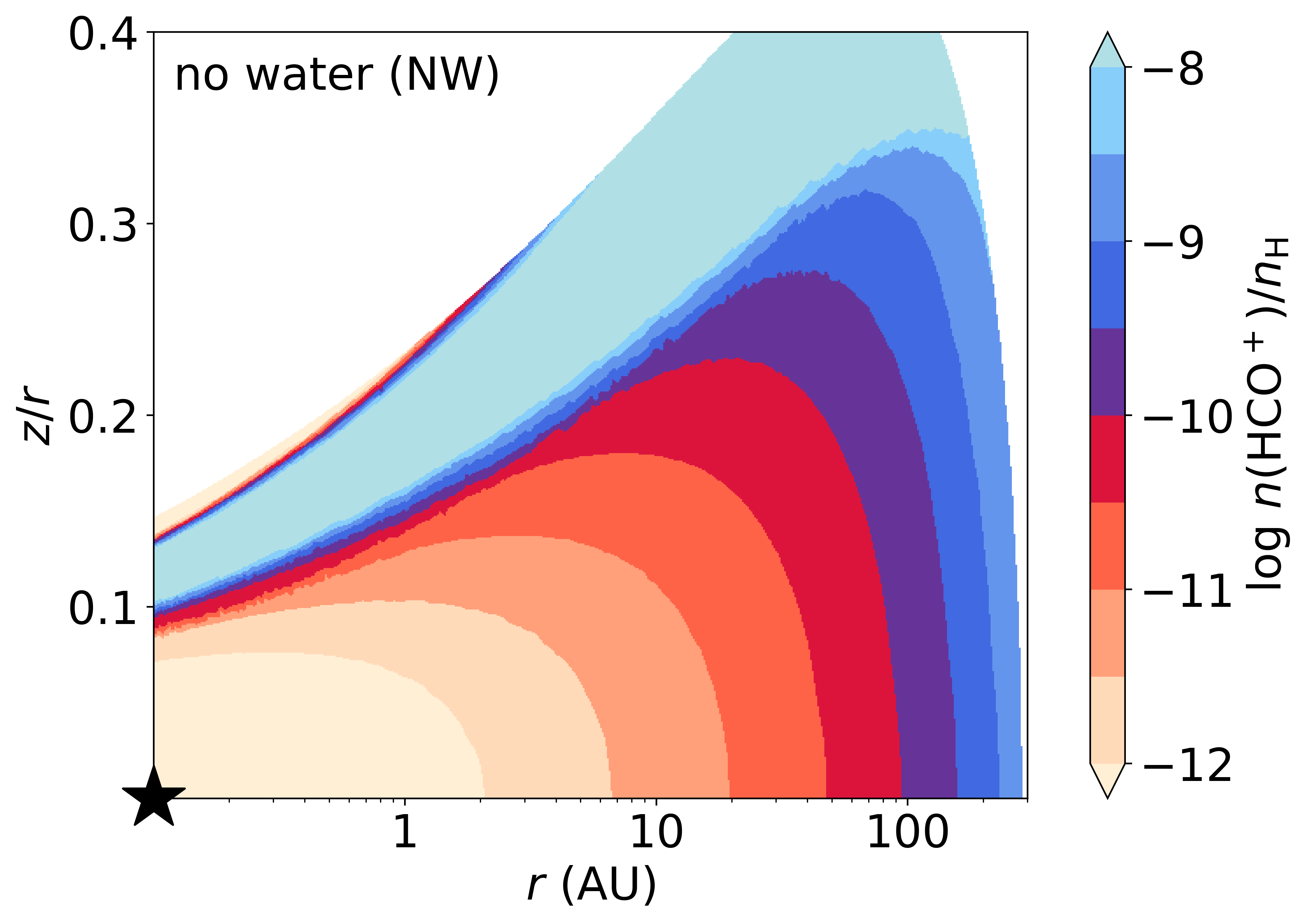}
   \end{subfigure} 
   
   \begin{subfigure}{1\columnwidth}
   \centering
   \includegraphics[width=1\linewidth]{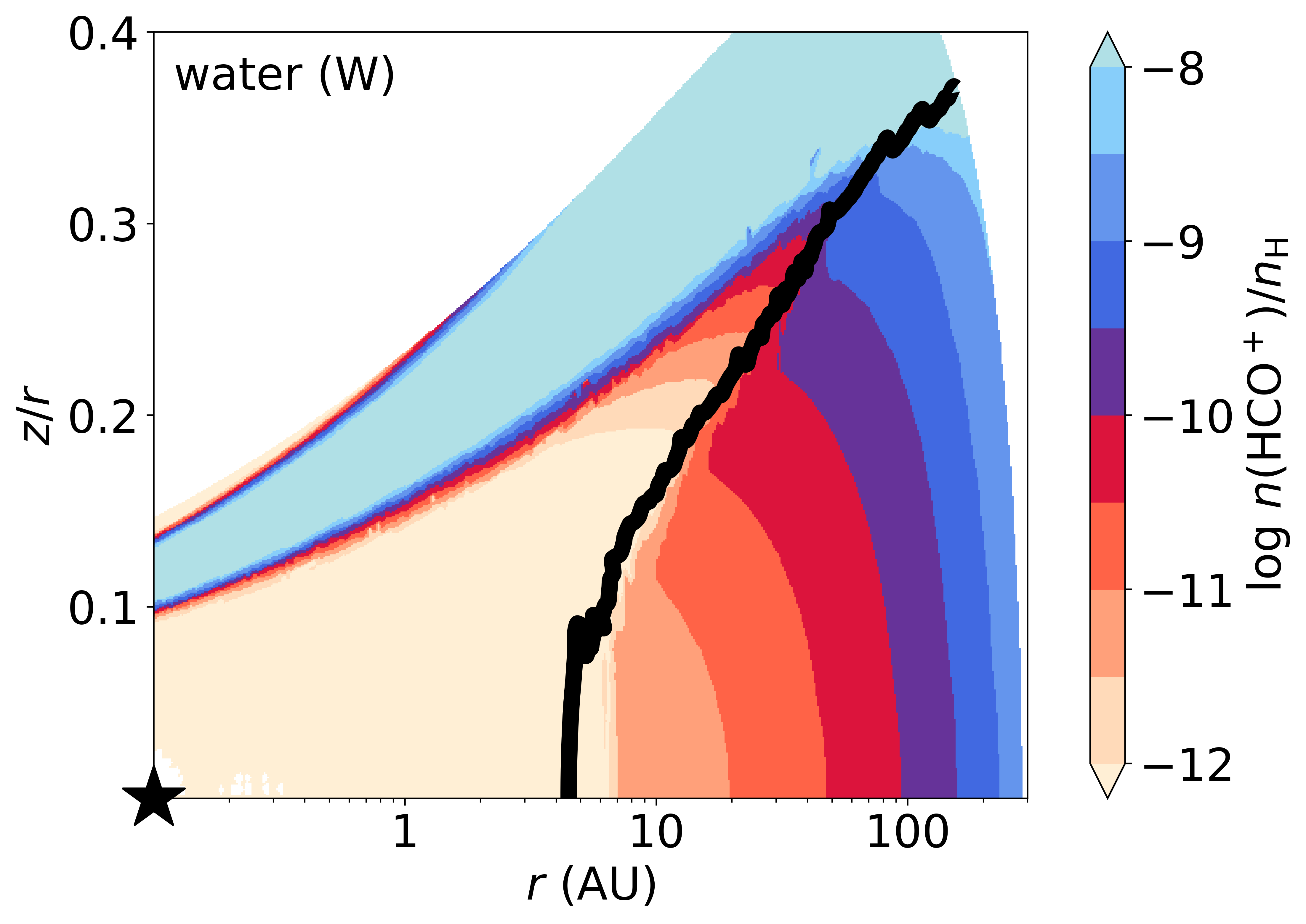}
   \end{subfigure}
   \caption{HCO$^+$ abundance calculated using chemical network NW (top) and network W (bottom). The water snow surface is marked with the black line in the bottom panel. The position of the star in indicated with the symbol of a black star.  }
   \label{fig:HCOpnetworkAD}
\end{figure}

The predicted abundances of HCO$^+$ by chemical networks NW and W are presented in Fig.~\ref{fig:HCOpnetworkAD}. The water snow surface in network W is computed as the surface where 50\% of the total water abundance is in the gas-phase and 50\% is frozen-out onto the grains and is indicated by the black line. In this model, the water snowline, the midplane location where water freezes-out, is located at 4.5~AU. Comparing the two chemical networks outside the water snow surface shows a great similarity in the predicted HCO$^+$ abundances. This is due to the fact that water is frozen out outside the water snow surface in network W, hence there is very little gas-phase water available for the destruction of HCO$^+$. In network W, the HCO$^+$ abundance drops inside the snowline, and is in general at least two orders of magnitude lower than in network NW. 
Up to 2~AU outside the water snowline HCO$^+$ is still efficiently destroyed by the small amount of water that is present in the gas-phase. Similar effects were found for N$_2$H$^+$ and CO \citep{Aikawa2015, vantHoff2017}. 
As HCO$^+$ is efficiently destroyed by gas-phase water, it thus is a good chemical tracer of the water snowline. 

The morphology of the HCO$^+$ distribution predicted by chemical network W is similar to the morphology predicted by full chemical networks \citep{Walsh2012, Walsh2013, Agundez2018}. Chemical network W agrees with the full chemical networks quantitatively inside the water snowline where all studies predict a low HCO$^+$ abundance of $\lesssim10^{-14}$. Moreover, the full networks and network W all show a jump of at least one order of magnitude in the HCO$^+$ abundance over a radial range of 5~AU outside the water snowline. 
In addition, these models predict a layer starting at $z/r \sim 0.1$ where the HCO$^+$ abundance reaches a high abundance of $10^{-6}-10^{-7}$ (\citealt{Walsh2012, Walsh2013} and \citealt{Agundez2018}). Yet, this layer is not expected to contribute much to the column density and emission of HCO$^+$, because the densities are low in this layer. 
The midplane abundances at 10~AU are $\sim10^{-12}$ in the full networks and $10^{-11}$ in network W, though the gradient in the HCO$^+$ abundance is very steep between the water snowline at 4.5~AU and a radius of 10~AU in network W complicating comparison. An HCO$^+$ abundance of $10^{-12}$ is reached at 7~AU. 
The HCO$^+$ abundance at 100~AU lies within the range of HCO$^+$ abundances predicted by the full chemical models. Our small network is thus suited to study HCO$^+$ as a tracer of the water snowline.

\subsubsection{Water vs. electrons}
HCO$^+$ only acts as a tracer of the water snowline if the destruction of HCO$^+$ by electrons is not dominant over the destruction by gas-phase water. The dissociative recombination of HCO$^+$ with an electron is included in both chemical networks. 
The rates of the reactions in Eq.~\ref{eq:HCOpH2O} and Eq.~\ref{eq:HCOpem}, $R_{\mathrm{H_2O}}$ and $R_{\mathrm{e^-}}$, respectively, are given by: 
\begin{align}
&R_{\mathrm{H_2O}} = k_{\mathrm{H_2O}} n(\mathrm{HCO^+}) n(\mathrm{H_2O})\ \text{and} \\
&R_{\mathrm{e^-}} = k_{\mathrm{e^-}} n(\mathrm{HCO^+})n(\mathrm{e^-}).
\end{align} 
Using Table \ref{tab:reactions}, the ratio of these reaction rates is given by:
\begin{align}
\frac{R_{\mathrm{e^-}}}{R_{\mathrm{H_2O}}} = 96 \left ( \frac{T_{\mathrm{gas}}}{300 \mathrm{K}} \right ) ^{-0.19} \frac{n(\mathrm{e^-})}{n(\mathrm{H_2O})}.
\end{align}
Therefore, gas-phase water is the dominant destruction mechanism of HCO$^+$ if the number density of gas-phase water is about two orders of magnitude larger than the number density of electrons.

\begin{figure}
   \centering
   \includegraphics[width=1\linewidth]{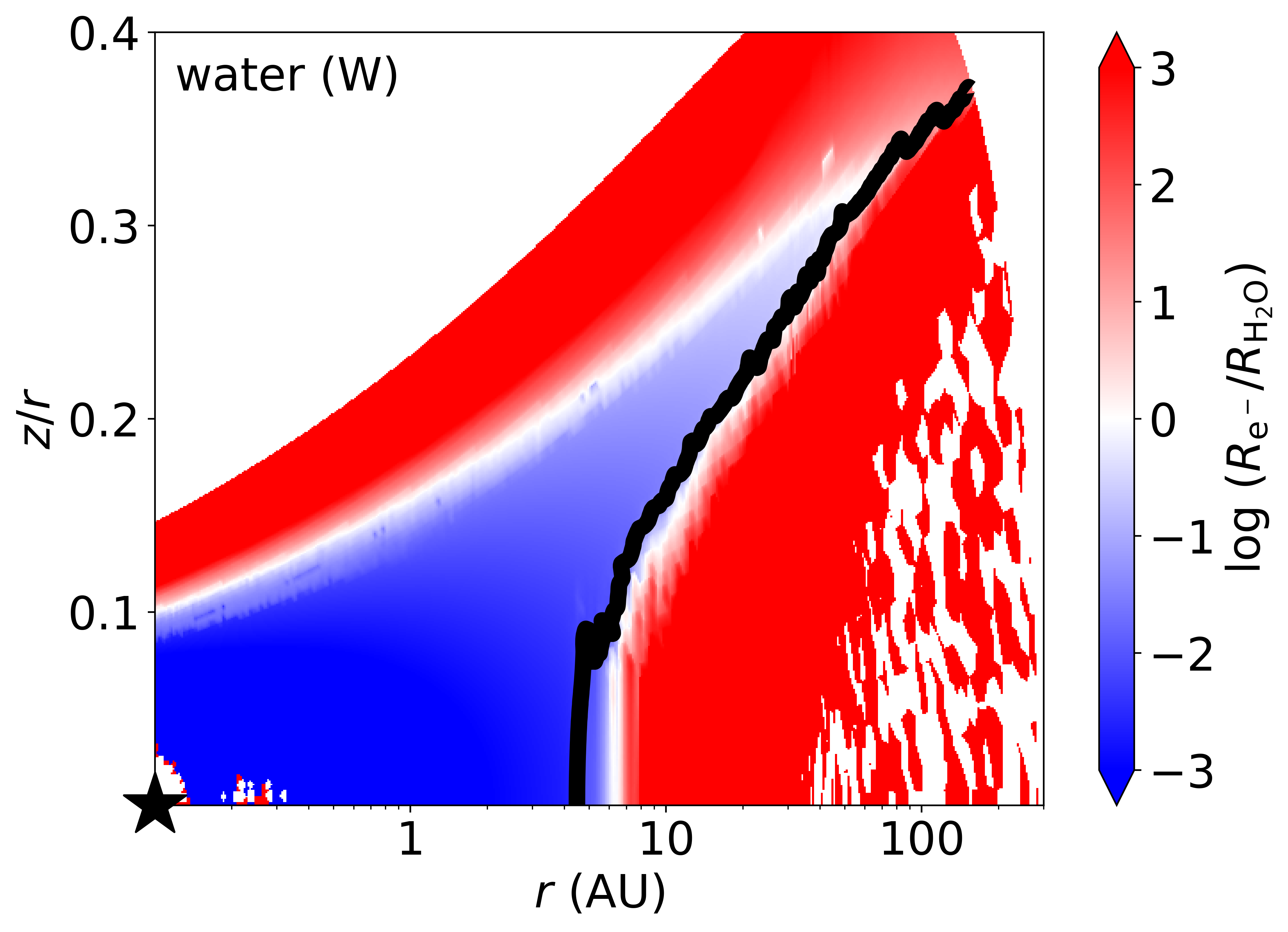}
      \caption{Relative reaction rates of HCO$^+$ destruction by electrons ($R_{\mathrm{e^-}}$) and water ($R_{\mathrm{H_2O}}$). Gas-phase water is the dominant destruction mechanism of HCO$^+$ in the blue regions and electrons are the dominant destruction mechanism in the red regions. The water snow surface is indicated by the black line.}
         \label{fig:xelectrons}
\end{figure}

Fig. \ref{fig:xelectrons} shows this ratio for chemical network W. The region where electrons are the dominant destruction mechanism ($R_{\mathrm{e^-}} > R_{\mathrm{H_2O}}$) is indicated in red and the region where water is dominant ($R_{\mathrm{e^-}} < R_{\mathrm{H_2O}}$) is indicated in blue. The first region exists mostly outside the water snow surface as water is frozen out. 
The latter region exists mostly inside and above the water snow surface, because there is plenty of water to destroy HCO$^+$ in these regions. Therefore HCO$^+$ is a good chemical tracer of the water snowline even though electrons are its dominant destruction mechanism in most of the disk.

\subsubsection{Effect of the CO, H$_2$O abundance, and cosmic ray ionization rate on the HCO$^+$ abundance} \label{sec:initial_cond_abu}
In this section we investigate the choice of initial conditions for chemical network W listed in Table~\ref{tab:params_chem_net}. The HCO$^+$ abundance outside the water snow surface in chemical network W can be approximated analytically (for details see Appendix~\ref{sec:appxHCOp}):
\begin{align}
x(\mathrm{HCO^+}) = \sqrt{\frac{\zeta_{\mathrm{c.r.}}}{k_{\mathrm{e^-}}n(\mathrm{H_2})}}. \label{eq:xHCOp}
\end{align}
Based on this equation, it is expected that the abundance of HCO$^+$ scales as the square root of the cosmic ray ionization rate in the disk region outside the water snowline. This analytical prediction is consistent with the predictions made by the numerical solution to chemical network W (top left panel in Fig.~\ref{fig:icchemnet}), where the HCO$^+$ column density decreases by one order of magnitude if the cosmic ray ionization rate decreases by two orders of magnitude. 

Furthermore, the HCO$^+$ abundance is expected to be independent of the initial CO and H$_2$O abundance outside the water snow surface based on the analytical approximation. 
This is in line with the predictions by the numerical solution to chemical network W (see HCO$^+$ column density in Fig.~\ref{fig:icchemnet}). The only major difference between the analytical approximation and the numerical solution occurs inside the water snowline. If the initial abundance of gas-phase water increases up to $\sim5\times 10^{-5}$ as expected for water ice in cold clouds, the column density of HCO$^+$ inside the water snowline decreases. This is because the higher abundance of gas-phase water destroys more HCO$^+$.  
 There is no significant effect on the HCO$^+$ column density outside the water snowline as all water is frozen out in that region of the disk. In summary, HCO$^+$ shows a steep jump in its column density around the water snowline for all initial conditions discussed in this section. Therefore, the results in the following sections do not depend critically on the choice of initial conditions. Further details on the initial conditions can be found in Appendix~\ref{sec:appinit}.

\subsection{HCO$^+$ and H$^{13}$CO$^+$ emission} \label{sec:results_rad_cut}
\begin{figure}
\centering
\includegraphics[width=1\linewidth]{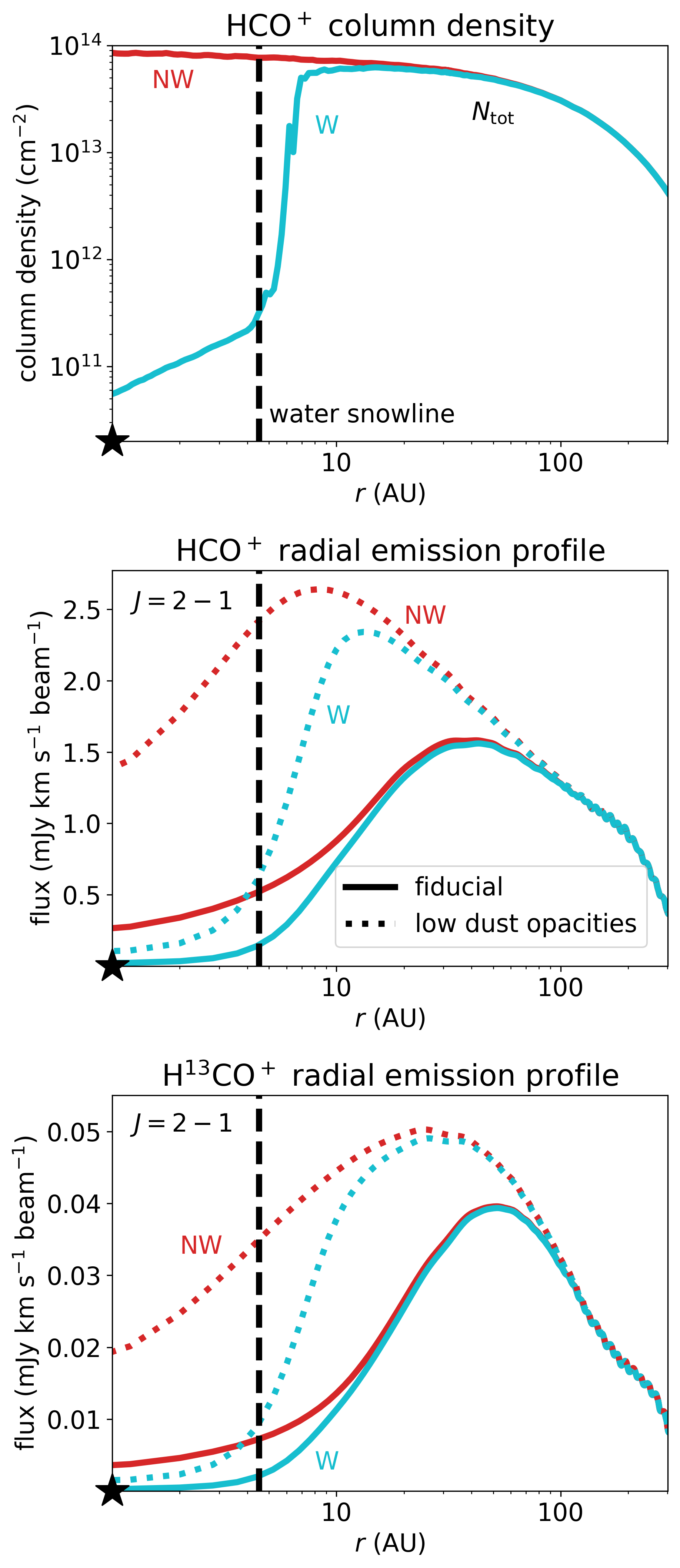}
\caption{Top panel: total HCO$^+$ column densities (top panel). Middle and bottom panel: HCO$^+$ $J=2-1$ radial emission profiles and H$^{13}$CO$^+$ $J=2-1$ radial emission profiles along the major axis predicted for chemical network NW (red) and chemical network W (light blue). The dotted lines refer to a model with very low dust opacities to lower the effects of continuum optical depth (see Section~\ref{sec:tau_cont}). The water snowline is indicated with the black dashed line. The radial emission profiles are convolved with a $0\farcs 05$ beam. }
\label{fig:radialmodelandobs}
\end{figure}

Observations do not trace the local abundances, only the emission which is closely related to the column densities if the emission is optically thin. 
The column densities of HCO$^+$ and the predicted radial emission profiles of HCO$^+$ and H$^{13}$CO$^+$ in network NW and W are shown in Fig.~\ref{fig:radialmodelandobs}. Similar to the abundance plots of HCO$^+$, the total HCO$^+$ column density jumps by a factor $\sim230$ over a radial range of 3~AU outside the water snowline in network W (solid blue line, top panel). Therefore, the HCO$^+$ column density shows a clear dependence on the presence of gas-phase water and the high HCO$^+$ abundance in the surface layers of the disk does not contribute much to the column density.

On the other hand, the shapes of the radial emission profiles of the HCO$^+$ $J=2-1$ transition are very similar for networks NW and W and the profiles peak at the same radius outside the water snowline (solid lines, middle panel). This is different from the HCO$^+$ column densities where only network W predicts the HCO$^+$ column density to peak outside the water snowline. The drop in the column density around the water snowline translates into a relative maximum difference of only 24\% between the radial emission profiles of the $J=2-1$ transition of HCO$^+$. This maximum relative difference is defined as the maximum difference between the flux predicted by network NW and W compared to the maximum flux predicted by network W and is typically located at the water snowline.

HCO$^+$ is more abundant than H$^{13}$CO$^+$, hence H$^{13}$CO$^+$ emission is expected to be optically thin, whereas the HCO$^+$ emission is optically thick. To test if this affects the ability to trace the water snowline, we predict emission from the less abundant isotopologue H$^{13}$CO$^+$. 
The radial emission profiles of H$^{13}$CO$^+$ for both chemical networks are shown in the bottom panel of  Fig.~\ref{fig:radialmodelandobs}. Qualitatively the profiles are similar to the predicted emission for HCO$^+$. Though the expected flux is about 39 times lower for H$^{13}$CO$^+$ than for HCO$^+$. 
This difference in flux is less than a factor 70, which indicates that the HCO$^+$ emission is indeed at least partially optically thick. The relative maximum difference between chemical network NW and W is 13\% for H$^{13}$CO$^+$, which is almost half of the corresponding number for HCO$^+$. So the effect of the water snowline is also seen in the emission of the optically thin H$^{13}$CO$^+$, but even less prominently. The relative difference for H$^{13}$CO$^+$ is smaller than for HCO$^+$ because H$^{13}$CO$^+$ emits from a lower layer in the disk due to its lower optical depth. Therefore more dust is present above the layer where H$^{13}$CO$^+$ emits, hence the effect of dust is larger for H$^{13}$CO$^+$ than for HCO$^+$.

These models show very similar radial emission profiles predicted for network NW and W and only a small difference in flux. Both networks predict ring shaped emission, regardless of the presence of the water snowline. This decrease in line flux towards the center of the disk can have different origins. The first one is a decrease in the abundance of the observed molecule, discussed in the previous section, which is the effect we aim to observe. However, the optical depth of the continuum emission and molecular excitation effects can change the emission as well, potentially obscuring the effect we would like to trace.  

\subsubsection{Continuum optical depth} \label{sec:tau_cont}
\begin{figure}
\centering
\includegraphics[width=1\linewidth]{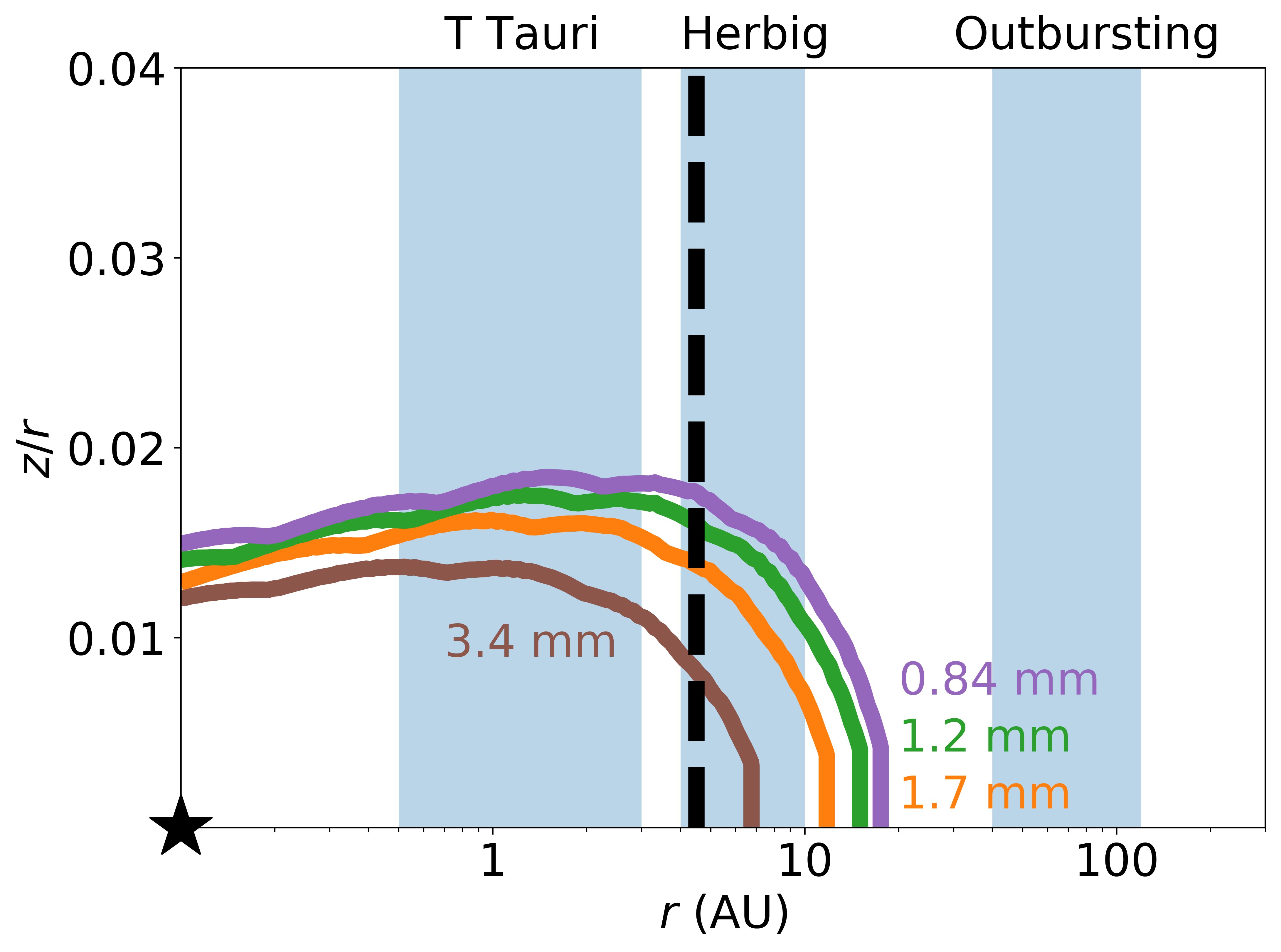}
\caption{$\tau_{\mathrm{dust}} = 1$ surface at the wavelength of the HCO$^+$ $J=1-0$ transition (3.4~mm, brown), $J=2-1$ (1.7~mm, orange), $J=3-2$ (1.2~mm, green) and $J=4-3$ (0.84~mm, purple). The radial range where the water snowline is expected for various types of sources is highlighted with the blue background. The water snowline in the model for a typical Herbig Ae disk is indicated with the back dashed line. }
\label{fig:taudust1surf}
\end{figure}

The continuum affects the line intensity when the optical depth of the continuum emission is larger than the optical depth of the line or when both are optically thick \citep{Isella2016}. To investigate the effect of the continuum emission, the $J=2-1$ transition of HCO$^+$ and H$^{13}$CO$^+$ was ray traced in a disk model where the dust opacities (dust mass absorption coefficients $\kappa_{\nu}$) are divided by a factor of $10^{10}$, to cancel the effect of dust on the HCO$^+$ emission compared to the typical Herbig~Ae disk model. The results are shown as the dotted lines in Fig.~\ref{fig:radialmodelandobs}. Comparing these radial emission profiles with the fiducial model shows that the effect of dust is small for network W in the inner $\sim$5~AU (solid versus dotted blue lines) even though the dust in the typical Herbig Ae disk model is optically thick out to $\sim$14~AU (see the orange line in Fig.~\ref{fig:taudust1surf}). The reason for this is that the HCO$^+$ abundance is low in the inner disk for network W. Therefore, little emission is expected, hence there is also little emission that can be affected by the dust. 

However, the dust can mimic the effect of gas-phase water, which greatly complicates matters. That the effect of the dust is significant, can be seen when the radial emission profiles predicted by network NW in the fiducial disk model and in the model with very low dust opacities (solid versus dotted red lines) are compared. This shows that the dust optical depth is largely responsible for the drop in emission in the center in network NW and for the fact that the radial emission profiles for network NW and W look very similar in the fiducial disk model. 
Therefore, to locate the water snowline using HCO$^+$, a disk with a high gas-to-dust mass ratio would be most suitable. However, recent work by \citet{Kama2020} has shown that disks with high gas-to-dust mass ratios around Herbig Ae stars are rare. Another option is to target HCO$^+$ in even warmer sources than Herbig~Ae stars. Fig.~\ref{fig:taudust1surf} shows that in more luminous sources, such as outbursting Herbig sources, the snowline is expected to shift to radii where the dust is optically thin at the wavelengths of the $J=1-0$ to $J=4-3$ transitions of HCO$^+$. Note however that the $\tau_{\mathrm{dust}} = 1$ surfaces not only depend on the frequency but also on the mass of the disk. Nonetheless, typical disk masses are not expected to be more than an order of magnitude higher as that would make them gravitationally unstable \citep{Booth2019, Booth2020, Kama2020}. 

\subsubsection{Molecular excitation}
\begin{figure}
\centering
\includegraphics[width=1\linewidth]{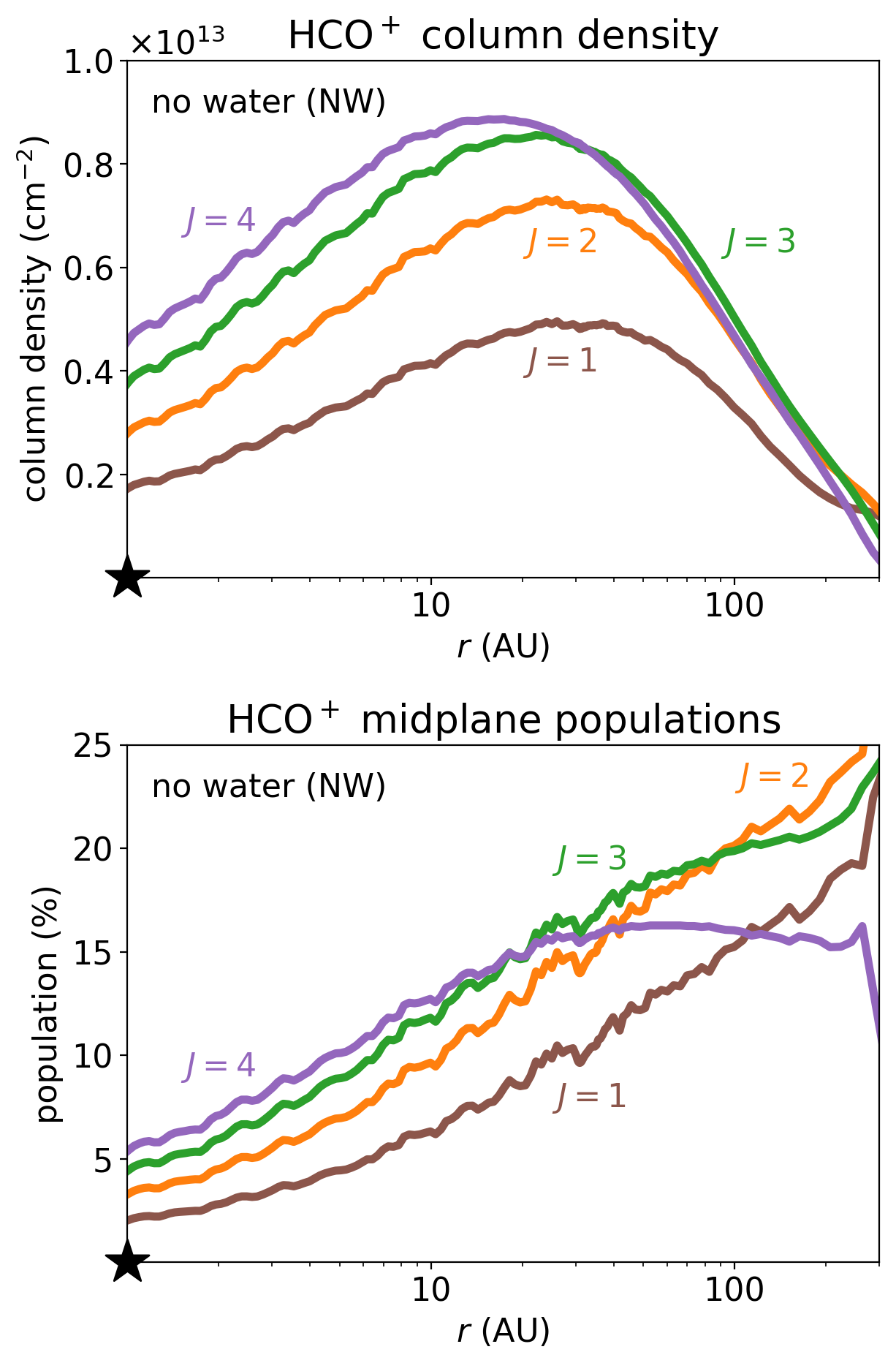}
\caption{HCO$^+$ column density (top) and midplane populations (bottom) for the $J=1$ (brown), $J=2$ (orange), $J=3$ (green), $J=4$ (purple) levels as a function of radius for chemical network NW in the typical Herbig disk model. }
\label{fig:pops}
\end{figure}

Decreasing the dust opacities by a factor of $10^{10}$ cannot fully explain the unexpected ring-shaped emission found by network NW. A second effect that contributes to the decrease in flux in the center is the temperature dependence of the $J$-level populations of HCO$^+$. The column densities and midplane populations of several $J$-levels of HCO$^+$ in chemical network NW are shown in Fig.~\ref{fig:pops}. 
The column density of the $J=2$ level of HCO$^+$ in chemical network NW peaks at a radius of $\sim$30~AU, while the total HCO$^+$ column density, $N_{\mathrm{tot}}$, peaks on-source (red line in top panel of Fig.~\ref{fig:radialmodelandobs}).

The radius where the column density of the $J=2$ level peaks roughly coincides with the radius where the emission of the $J=2-1$ transition of HCO$^+$ associated with network NW with low dust opacities starts to drop. 
Therefore, both the optical depth of the continuum emission as well as the temperature dependence of the $J$-level population contribute to the drop in emission seen in the center.

\subsubsection{Chemistry}
Finally there could also be a chemical effect in network NW as the midplane abundance of HCO$^+$ decreases towards the center of the disk. Yet, the total column density of HCO$^+$ keeps increasing up to a radius of 0.07~AU, which is too small to explain the decrease in flux out to $\sim$50~AU seen in the radial emission profile associated with network NW in the fiducial model or to $\sim$10-25~AU in the model with the low dust opacities. 
Still the drop in the emission associated with network W is for a small part due to a decrease in the total HCO$^+$ column density.

\subsubsection{What causes HCO$^+$ rings?}
In summary, all three effects contribute to the decrease in flux in the inner regions of the disk. Observing different transitions of HCO$^+$ and H$^{13}$CO$^+$ will give a different balance of these effects.
Fig.~\ref{fig:pops} shows that higher $J$-levels are more populated in the inner disk because the inner disk is too warm and dense to highly populate the $J=2$ level of HCO$^+$. However, the emitting frequency of HCO$^+$ increases as higher transitions are observed, so also the continuum optical depth increases.

Therefore it is difficult, yet crucial, to disentangle the effects of the water snowline and of the population of HCO$^+$ and the optical depth of the continuum emission. 
The effect of the continuum optical depth can be quantified by observing another molecule that emits from the same region as HCO$^+$ or H$^{13}$CO$^+$ but does not decrease in column density towards the star such as a CO isotopologue. The excitation and column density of HCO$^+$ need to be inferred from detailed modelling. 
In conclusion, the results in this section show that even though the HCO$^+$ abundance changes by at least two orders of magnitude around the water snow surface, it is observationally complicated to verify this.

\subsection{Line profiles} \label{sec:results_line_profile}
\begin{figure}
\centering
\includegraphics[width=1\linewidth]{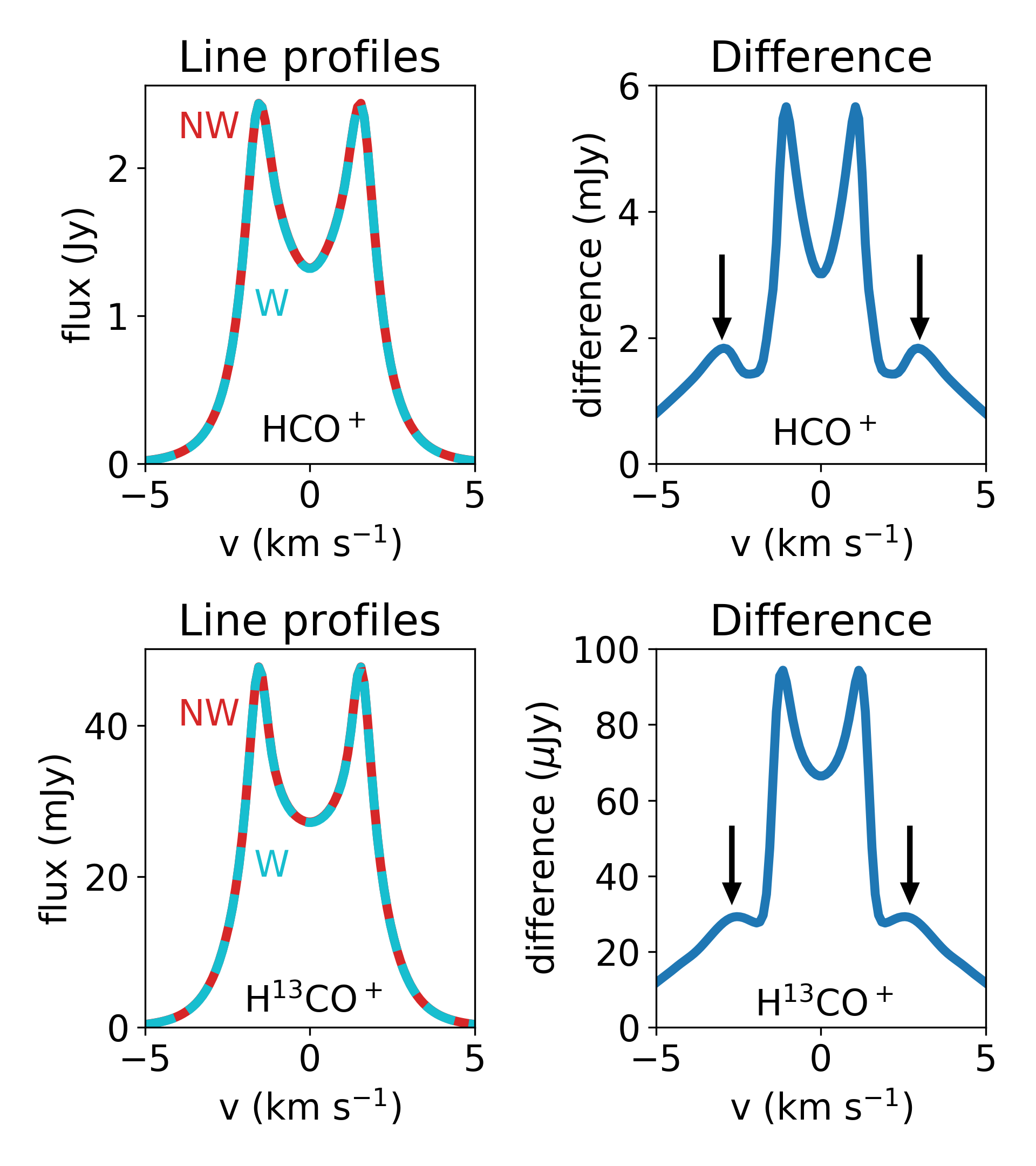}
\caption{Left panels: line profile for the $J=2-1$ transition of HCO$^+$ (top) and H$^{13}$CO$^+$ (bottom) for chemical network NW (red) and W (light blue). Right panels: difference between the line profiles in the left column. The black arrows indicate where the effect of the water snowline is expected. Note the differences in the vertical axes. }
\label{fig:lineprofile}
\end{figure}

The discussion above has shown that locating the water snowline using the HCO$^+$ radial emission profile is difficult. The velocity resolved line profile provides an alternative method as the removal of HCO$^+$ in the inner disk is expected to remove flux at the highest Keplerian velocities. Accordingly, the line profile predicted by chemical network W is expected to be narrower than the line profile predicted by network NW. 

The line profiles predicted by the two chemical networks for the $J=2-1$ transition of HCO$^+$ and H$^{13}$CO$^+$ are shown in left column of Fig.~\ref{fig:lineprofile}. To show the difference between the models we subtracted the spectrum associated with network W from the spectrum associated with network NW. The result is shown in the  right column of Fig.~\ref{fig:lineprofile}. This column shows that even though the line profiles look almost identical in the left column, there is an additional bump in the difference between them in the line wings (indicated with the black arrows). However, this difference is only 1.75~mJy for HCO$^+$ and 28~$\mu$Jy for H$^{13}$CO$^+$ i.e. 0.6\% and 0.3\% of emission, which cannot be detected with a reasonable signal-to-noise ratio. Moreover, detailed comparison with the line profile associated with a molecule that does not show a decrease in the column density in the inner disk would be needed as only the flux associated with network W would be observed and not the flux associated with network NW. Hence, it is very difficult to use the line profiles to locate the water snowline in disks around Herbig Ae stars.

\section{Comparison with observations} \label{sec:obs}
The best targets to observe the water snowline using HCO$^+$ are warm disks, because these disks have a water snowline at a large enough radius for ALMA to resolve. In addition, disks without deep gaps in the gas and dust surface density around the expected location of the snowline are preferred, as these gaps complicate the interpretation of the HCO$^+$ emission. 

One of the most promising targets to locate the water snowline in a disk is the young outbursting star V883 Ori, with a current luminosity of $\sim$218~L$_{\odot}$ \citep{Furlan2016}. During the outburst, the luminosity is greatly increased, which heats the disk, shifting the water snowline outwards compared to regular T~Tauri and even Herbig Ae disks. This not only allows for observations with a larger beam, it also mitigates the effects of the dust as the optical depth of the continuum emission decreases with radius. Previous observations of V883 Ori have inferred the location of the water snowline at 42~AU based on a change in the dust emission \citep{Cieza2016}. An extended hot inner region is consistent with the detection of many complex organic molecules \citep{Lee2019}. However, an abrupt change in the continuum optical depth or the spectral index are not necessarily due to a snowline, as shown for the cases of CO$_2$, CO and N$_2$ \citep{Huang2018DSHARPII, Long2018, vanTerwisga2018}. Methanol observations suggest that the water snowline may be located at a much larger radius of $\sim$100~AU if the methanol observations are optically thin and most of its emission originates from inside the water snowline \citep{vantHoff2018methanol}. Observations of HCO$^+$ have the potential to resolve this discrepancy.

\subsection{H$^{13}$CO$^+$ observations}
A promising dataset to locate the water snowline in V883 Ori is the band~7 observation by \citet{Lee2019} (project code: 2017.1.01066.T, PI: Jeong-Eun Lee), which contains the H$^{13}$CO$^+$ $J=4-3$ line. To date, these observations  are the only H$^{13}$CO$^+$ observations in a protoplanetary disk, to our knowledge, at sufficient spatial and spectral resolution with sufficient sensitivity to potentially resolve the water snowline. 
The synthesized beam of these observations is $0\farcs 2$ (40~AU radius, 80~AU diameter at a distance of $\sim$400~pc; \citealt{Kounkel2017}), and the spectral resolution is 0.25~km~s$^{-1}$. The continuum is created using all line-free channels, which are carefully selected to exclude any line emission. The line data are continuum subtracted using these continuum solutions. Using the CASA 5.1.1 {\it tclean} task with a Briggs weighting of 0.5, line images are made with a mask of about 2" in diameter centered on the peak of the continuum.

\begin{figure}
\centering
\includegraphics[width=1\linewidth]{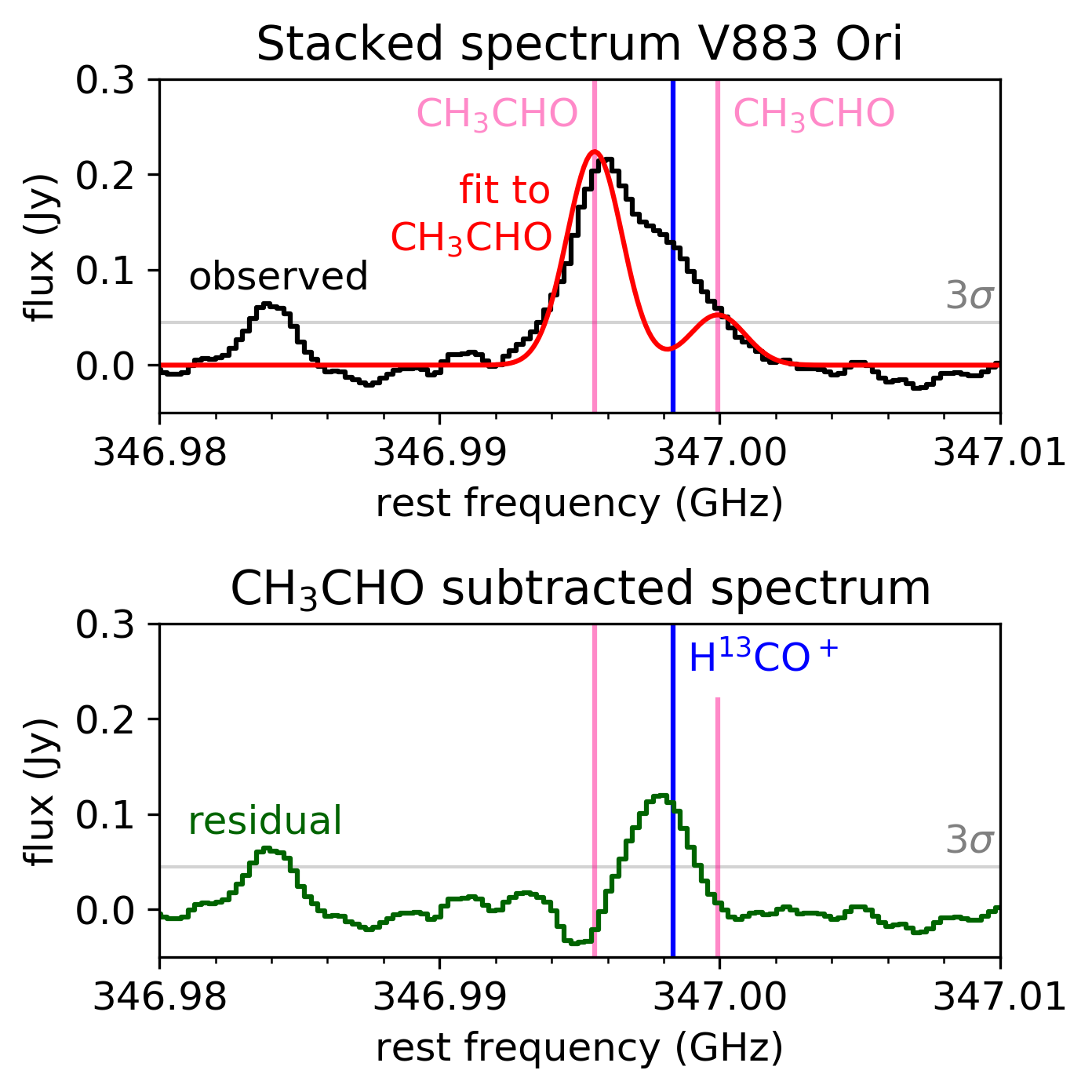}
\caption{Top panel: observed and stacked spectrum of V883 Ori (black) and a fit of the CH$_3$CHO emission (red); bottom panel: spectrum where the CH$_3$CHO emission is subtracted. The H$^{13}$CO$^+$ $J=4-3$ emission line (vertical dark blue line) is blended with the CH$_3$CHO lines (vertical pink lines). The 3$\sigma$ noise level is indicated with the horizontal grey line. }
\label{fig:V883Orispec}
\end{figure}

Following the approach of \citet{Lee2019}, the line observations are velocity-stacked using a stellar mass, $M_{\star}$ = 1.3~M$_{\odot}$, an inclination of 38$\degree$, and a position angle of 32$\degree$ \citep{Cieza2016}. This technique reduces line blending because it makes use of the Keplerian velocity of the disk to calculate the Doppler shift of the emission in each pixel \citep{Yen2016, Yen2018}. This Doppler shift is then used to shift the spectrum of each pixel to the velocity of the star, before adding the spectra of all pixels. The noise level of $\sim$15~mJy per spectral bin is determined using an empty region in the stacked image.

The observed spectrum of the inner $0\farcs 6$ after line stacking is shown as the black line in the top panel of Fig.~\ref{fig:V883Orispec}.
The rest frequency of the H$^{13}$CO$^+$ $J=4-3$ transition lies between the rest frequencies of the 18$_{7, 12}$~-~17$_{7, 11}$, E transition of acetaldehyde (CH$_3$CHO) ($v=0$) at 346.9955~GHz and the two superimposed 18$_{7, 11}$~-~17$_{7, 10}$, E and 18$_{7, 12}$~-~17$_{7, 11}$, E transitions of CH$_3$CHO ($v_{\mathrm{T}}=2$) at 346.99991~GHz and 346.99994~GHz, respectively (Jet Propulsion Laboratory (JPL) molecular database; \citealt{Pickett1998}). In between these acetaldehyde transitions, a clear excess in emission is visible due to the $J=4-3$ transition of H$^{13}$CO$^+$. 
To quantify this excess emission, two Gaussian profiles are fitted with \texttt{curve\_fit}\footnote{The \texttt{curve\_fit} function is part of the Scipy package in python.} to model the acetaldehyde emission. To reduce the number of free parameters in the fit, the line frequencies are fixed to their respective rest frequencies listed above and the widths of the lines are fixed to 2~km~s$^{-1}$ following \citet{Lee2019}. In addition, emission within 1~km~s$^{-1}$ from the rest frequency of the $J=4-3$ transition of H$^{13}$CO$^+$ is excluded from the fit. 
The result is shown as the red line in the top panel of Fig.~\ref{fig:V883Orispec}. Subtracting this fit to the acetaldehyde emission from the observed spectrum clearly reveals emission of the $J=4-3$ transition of H$^{13}$CO$^+$. Therefore we conclude that H$^{13}$CO$^+$ is detected at $>$5$\sigma$ but is blended with lines that are expected to peak on-source, preventing a clean image.

The line blending of H$^{13}$CO$^+$ $J=4-3$ with acetaldehyde prevents us from using the only available dataset covering a transition of H$^{13}$CO$^+$ at sufficient spatial and spectral resolution and sensitivity to resolve the water snowline. The $J=3-2$ transition of H$^{13}$CO$^+$ is likely blended with CH$_3$OCHO as both of them are bright in the Class 0 source B1-c \citep{vanGelder2020}. However, the H$^{13}$CO$^+$ $J=2-1$ transition in the same Class 0 B1-c source is free from line blending (van 't Hoff et al. in prep.) and would therefore be the best line to target in sources with bright emission from complex organics. 

\subsection{HCO$^+$ observations} \label{sec:band6}

Another promising dataset for HCO$^+$ is a recent band~6 dataset (project code 2018.1.01131.S, PI: D.~Ru{\'\i}z-Rodr{\'\i}guez; Ru{\'\i}z-Rodr{\'\i}guez et al. in prep.). 
The $J=3-2$ transition of HCO$^+$ is detected with a high signal-to-noise ratio. The spatial resolution of $\sim0 \farcs 5$  
is insufficient to resolve the water snowline if it is located within a radius of $\sim90$~AU (180~AU diameter).

\begin{figure}
   \centering
  \begin{subfigure}{0.99\columnwidth}
  \centering
  \includegraphics[width=1\linewidth]{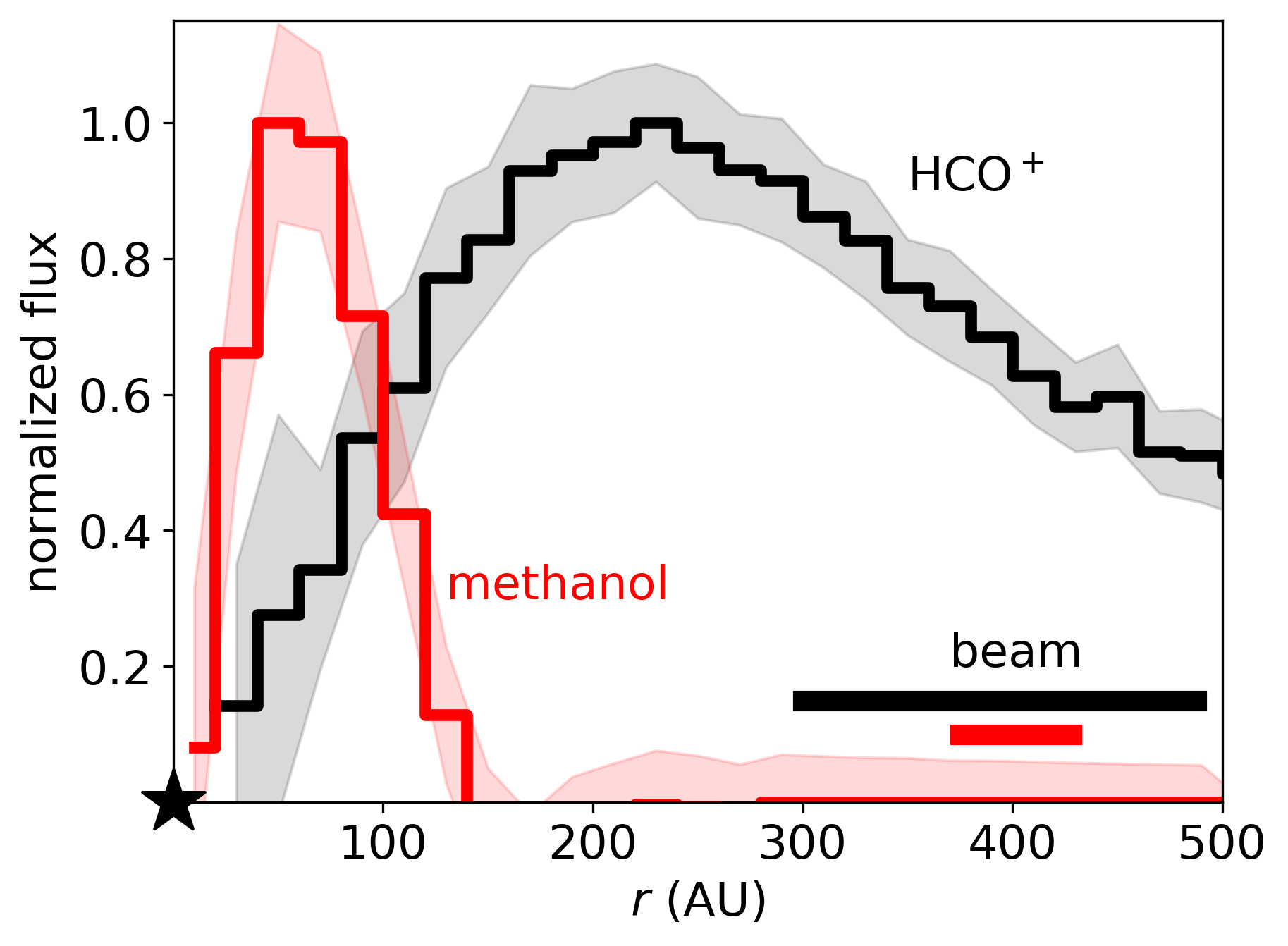}
\end{subfigure}%

\begin{subfigure}{0.99\columnwidth}
  \centering
    \includegraphics[width=1\linewidth]{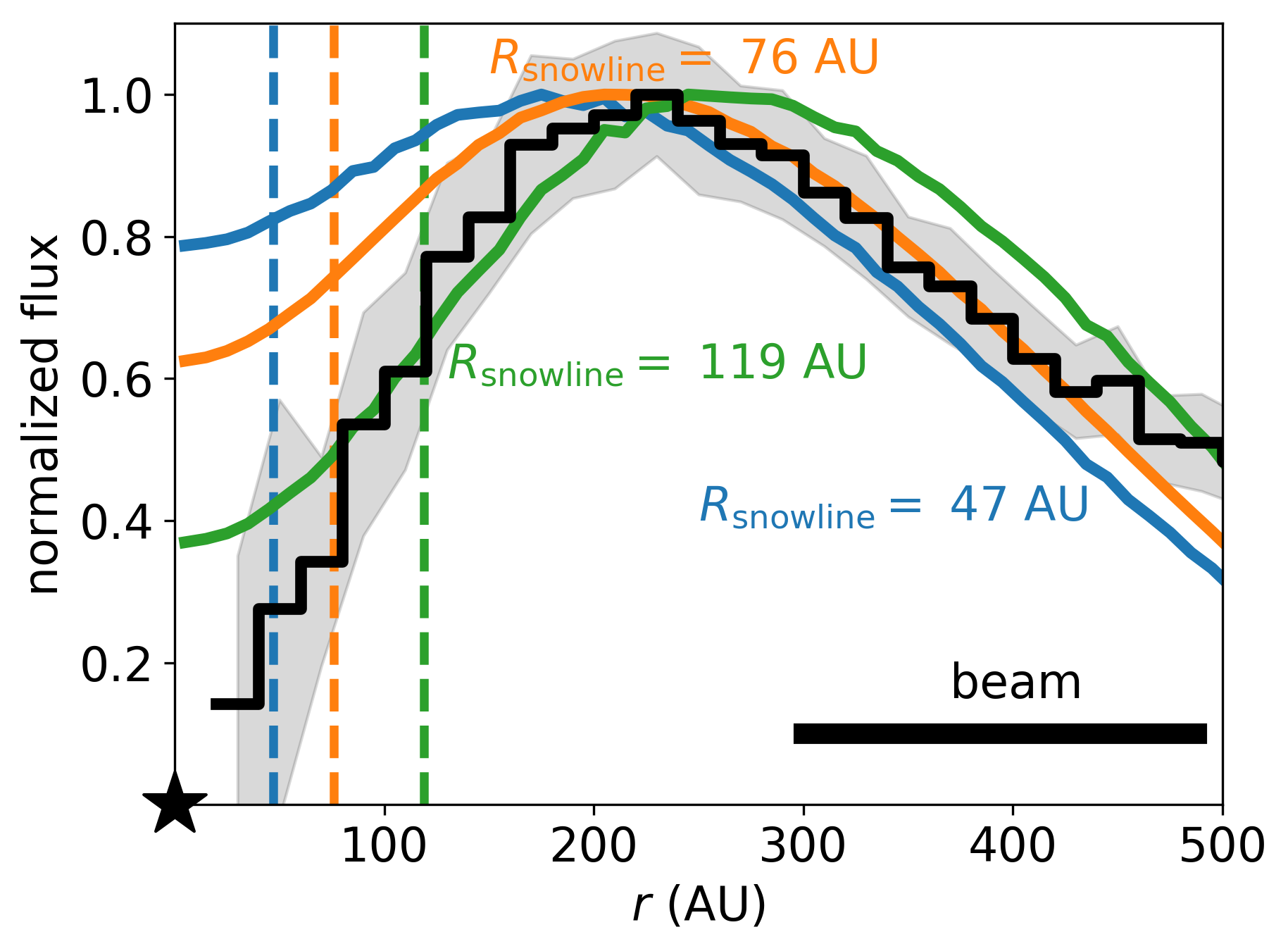}
\end{subfigure}

      \caption{Top: normalized azimuthal average of the observed HCO$^+$ $J=3-2$ flux in V883 Ori (black; Ru{\'\i}z-Rodr{\'\i}guez et al. in prep. and this work) and methanol 18$_3$-17$_4$ flux (red; \citealt{vantHoff2018methanol}). 
      Bottom: normalized azimuthal average of the observed HCO$^+$ flux (black; as in top panel) and modelled HCO$^+$ flux for a snowline at 47~AU (blue), 76~AU (orange) and 119~AU (green). The snowline locations of the models are indicated with dashed lines in corresponding colors. The position of the star is indicated with a black star and the beam is indicated by the black bar in the bottom right corner. }
         \label{fig:compareHCO+32_models_obs}
   \end{figure}

An image of the HCO$^+$ emission will be presented in Ru{\'\i}z-Rodr{\'\i}guez et al. (in prep.).
Here, a normalized, deprojected azimuthal average of the observed emission from the product data of the $J=3-2$ transition of HCO$^+$ is presented in the top panel of Fig.~\ref{fig:compareHCO+32_models_obs}. This azimuthal average shows that the HCO$^+$ emission is ring shaped. Comparing the band~6 HCO$^+$ emission with the band~7 emission from complex organic molecules presented in \citet{vantHoff2018methanol} and \citet{Lee2019}
, reveals that the emission from complex organic molecules such as methanol, 13-methanol, acetaldehyde and  H$_2$C$^{18}$O peak inside the HCO$^+$ ring and in some cases is even centrally peaked. This is a strong indication that the lack of HCO$^+$ emission in the center is not only due to the optical depth of the continuum and that HCO$^+$ is indeed tracing the water snowline in V883 Ori. Such an anti-correlation between HCO$^+$ and methanol has also been seen in protostellar envelopes \citep{Jorgensen2013}.

\subsection{Model HCO$^+$ images: locating the snowline} \label{sec:V883Ori_models}

\begin{figure*}
   \centering
   \includegraphics[width=1\linewidth]{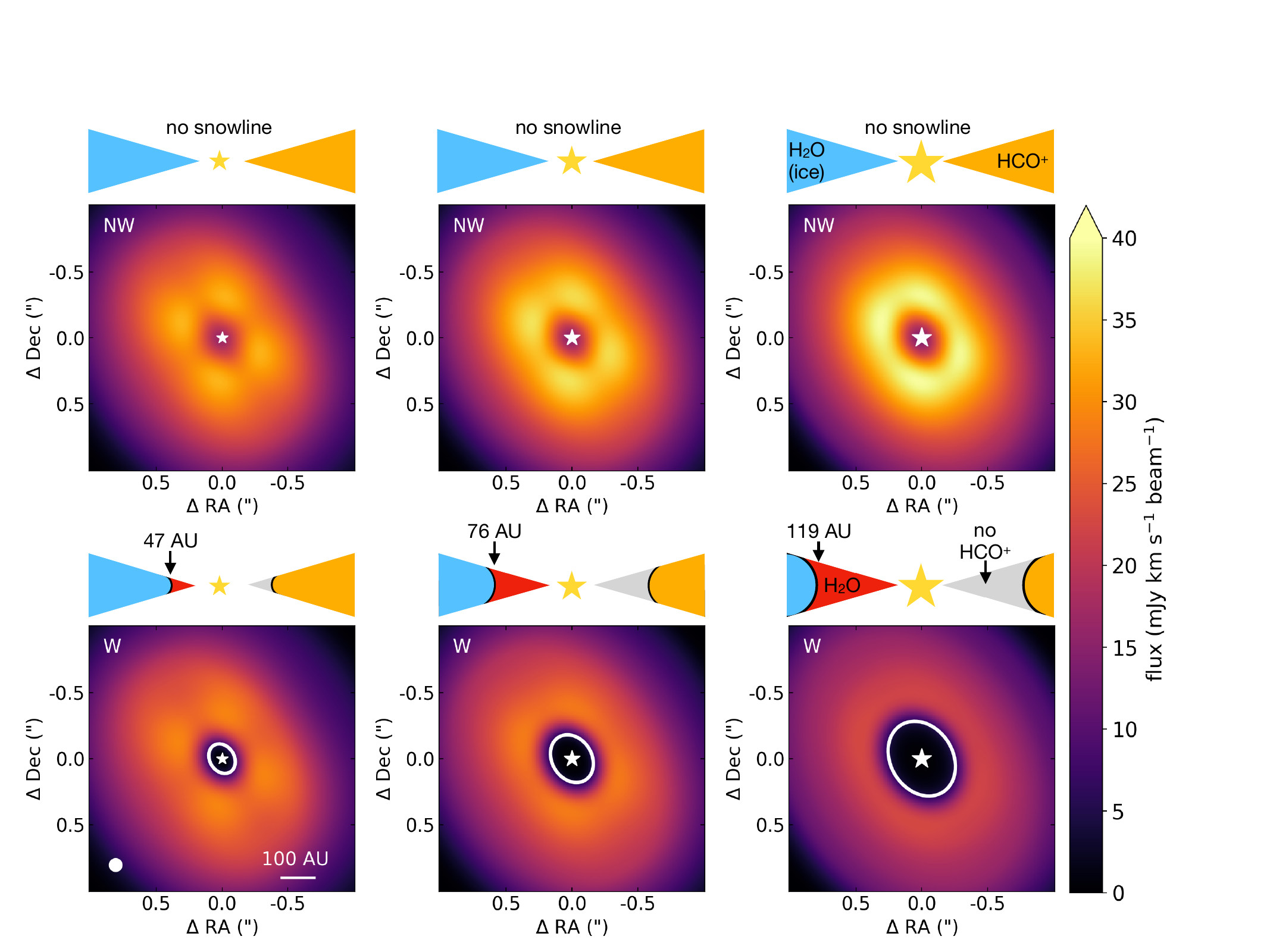}
      \caption{Integrated intensity maps for the $J=3-2$ transition of HCO$^+$ predicted by network NW (top row) and network W (bottom row). The snowline is indicated with a white ellipse and is located at 47 AU (left column), 76 AU (middle column) and 119 AU (right column). The cartoons above the individual panels provide a sketch of the model, where blue indicates water ice, red indicates gas-phase water, orange indicates a high abundance of HCO$^+$ and gray indicates a low abundance of HCO$^+$. The position of the star is marked with a yellow star in each cartoon and with white star in each panel. Note that the size of the star increases with increasing luminosity of the star. The $0 \farcs 1$ beam and a scale bar are indicated in the bottom left panel.}
         \label{fig:V883Ori_mom0}
\end{figure*}   

To quantify the location of the water snowline, we model the HCO$^+$ and H$^{13}$CO$^+$ emission in a representative model for V883 Ori. 
This model reproduces the previously observed flux of the $J=2-1$ transition of C$^{18}$O \citep{vantHoff2018methanol}, the HCO$^+$ and H$^{13}$CO$^+$ total flux discussed in the previous Sections, and mm continuum fluxes within a factor of $\sim$2. The \texttt{DALI} model parameters are presented in Table~\ref{tab:paramsdaliV883}. The main changes compared to the model for the typical Herbig Ae disk include the mass and radius of the disk and the mass and luminosity of the star. The characteristic disk radius is set to 75~AU to match the radial extent of emission from the $J=2-1$ transition of C$^{18}$O presented by \citet{vantHoff2018methanol}. The disk mass is estimated using the 9.1~mm continuum observations of the VANDAM survey \citep{Tobin2020} and the relation between the continuum flux and the disk mass \citep{Hildebrand1983}:
\begin{align}
M = \frac{D^2 F_{\nu}}{\kappa_{\nu}B_{\nu}(T_{\mathrm{dust}})},
\end{align}
with $D$ the distance to V883 Ori, $F_{\nu}$ the observed continuum flux, $\kappa_{\nu}$ the opacity and $B_{\nu}(T_{\mathrm{dust}})$ the Planck function for a dust temperature $T_{\mathrm{dust}}$. As V883 Ori is an outbursting source, a dust temperature of 50~K is assumed. Following \citet{Tychoniec2020}, a dust opacity of 0.28~cm$^2$g$^{-1}$ at a wavelength of 9.1~mm is used \citep{Woitke2016}. This gives an estimated disk mass of 0.25~M$_{\odot}$. 

The snowline has been estimated at 42~AU from dust \citep{Cieza2016}, but can be as far out as 100~AU based on CH$_3$OH \citep{vantHoff2018methanol}. Moreover, the outburst likely began before 1888 \citep{Pickering1890}. Approximately 25~years ago the luminosity was measured to be $\sim$200~L$_{\odot}$ higher than the current luminosity \citep{Strom1993, Sandell2001, Furlan2016} and it could have been much higher in the past. With a freeze-out time scale of 100-1000~yrs, the snowline may thus not be at the location expected from the current luminosity \citep{Jorgensen2013, Visser2015, Hsieh2019}. We therefore use three luminosities of $2\times 10^3$, $6\times 10^3$ and $1.4\times 10^4$~L$_{\odot}$ in our models, which put the snowline at 47, 76, and 119~AU, respectively.

The integrated intensity maps of the $J=3-2$ transition of HCO$^+$ for these models are shown in Fig.~\ref{fig:V883Ori_mom0}. The corresponding figures for the weaker HCO$^+$ and H$^{13}$CO$^+$ $J=2-1$ transitions are presented in Fig.~\ref{fig:V883Ori_mom0HCO+21} and Fig.~\ref{fig:V883Ori_mom0HiCO+21}. 
These lines are not expected to be contaminated by emission from complex organic molecules. The results in this section are convolved with a small $0 \farcs 1$ beam to make predictions for future high resolution observations.

The top row of Fig.~\ref{fig:V883Ori_mom0} shows the expected emission for network NW for the three different luminosities. The total HCO$^+$ emission does not depend strongly on the luminosity because the population of the $J=3$ level decreases with luminosity but the increase in the temperature of the emitting region cancels this effect as the HCO$^+$ emission is marginally optically thick.
Similar to the models for a disk around a typical Herbig Ae star, the moment 0 maps of network NW show ring shaped emission despite the fact that there is no snowline present in these models. The lack of emission in the center is dominated by absorption by dust. On top of that, the column density of the $J=3$ level of HCO$^+$ decreases in the inner parts of the disk. These two points are difficult to disentangle because the continuum optical depth is frequency and hence $J$-level dependent. However, most importantly, the location of the HCO$^+$ ring does not change as a function of luminosity in network NW.

This is different in the bottom row of Fig.~\ref{fig:V883Ori_mom0} where emission associated with network W is shown. The water snowline is indicated with a white ellipse and shifts outwards as the luminosity of the star increases. This is also reflected in the HCO$^+$ emission as the ring of HCO$^+$ shifts outwards together with the water snowline. Therefore, the location of the HCO$^+$ ring can be used to locate the water snowline provided that another molecule is observed to prove that the decrease in the HCO$^+$ flux in the center is not solely due to molecular excitation effects and absorption by dust. The excitation effect needs to be inferred from disk modelling. The effect of the optical depth of the continuum emission can be estimated by observing another molecule, that emits from the same disk region as HCO$^+$ and whose column density does not drop in the inner disk. The most obvious molecule for this purpose is C$^{18}$O. However, the HCO$^+$ $J=3-2$ emission comes from a layer closer to the midplane than the C$^{18}$O $J=2-1$ and $J=3-2$ emission. Therefore, the effect of the dust on the HCO$^+$ emission cannot be fully traced by the $J=2-1$ or $J=3-2$ transition of C$^{18}$O. More rare isotopologues such as C$^{17}$O or $^{13}$C$^{18}$O are more suited for this purpose, as well as other molecules such as complex organics as discussed in Section~\ref{sec:band6}.

The model predictions for the observed flux of the $J=3-2$ transition of HCO$^+$ in V883 Ori are compared with the observed flux in the bottom panel of Fig.~\ref{fig:compareHCO+32_models_obs}. The model results are convolved to the same spatial resolution as the observations before calculating the deprojected azimuthal average. Even though the beam is too large to resolve the snowline if it is located inside $\sim90$~AU, this figure clearly shows that the observed HCO$^+$ flux drops off steeper in the inner parts than that predicted by the model with a snowline at 47~AU. In addition, the HCO$^+$ ring shifts outwards with increasing snowline location. The observed peak location and gap depth in the center best match with a snowline between 76 and 119~AU.

A central cavity (approx. 1~beam in diameter at $0\farcs 35 \times 0 \farcs 27$ resolution corresponding to $\sim$60~AU resolution in radius) was also observed for CO $J=2-1$ and was attributed to the continuum optical depth by \citet{RuizRodriguez2017}. However, it is unlikely that the cavity observed in HCO$^+$ is solely due to the continuum as the emission from complex organic molecules peaks inside the HCO$^+$ ring ($0\farcs 23 \times 0\farcs 17$; \citealt{Lee2019}). Similarly, the high resolution methanol emission ($0\farcs 13\times 0\farcs 14$; \citealt{vantHoff2018methanol}) and the continuum emission ($0\farcs 03$; \citealt{Cieza2016}) peak well inside the HCO$^+$ cavity (Fig.~\ref{fig:compareHCO+32_models_obs}). Furthermore, the continuum becomes optically thin well within 100~AU in all three of our models (Figs.~\ref{fig:compareHCO+32_models_obs} and \ref{fig:V883Ori_mom0}), so the difference between the models is due to different snowline locations. Therefore, our analysis is consistent with a water snowline at $\sim$75-120~AU in V883 Ori and thus suggest that the sudden change in continuum opacity at 42~AU is uncorrelated with the water snowline.

One effect that needs to be taken into account in outbursting sources like V883 Ori is viscous heating. This effect heats the disk midplane, hence the water snowline could be at a larger radius than expected based on radiative heating alone. The $J=3-2$ transition of HCO$^+$ is marginally optically thick in our models. Targeting a low $J$ transition of HCO$^+$ or targeting the optically thin H$^{13}$CO$^+$, will allow to directly trace the midplane. Another effect that could occur due to viscous heating is self-absorption of HCO$^+$ as the surface layers of the disk could have a lower temperature than the viscously heated midplane. This could be resolved by observing multiple transitions of HCO$^+$ as higher $J$ lines are more optically thick and hence trace a layer higher in the disk. 

Taken together, our analysis of the observations suggests that HCO$^+$ is tracing the water snowline in V883 Ori and that the snowline is located outside the radius where the change in the continuum opacity is observed. 
Therefore, V883 Ori is the first example that shows that a change in the dust continuum opacity is not necessarily related to the water snowline, similar to what has been shown for the cases of CO, CO$_2$ and N$_2$ snowlines \citep{Huang2018DSHARPII, Long2018, vanTerwisga2018}.


\section{Conclusions} \label{sec:concl}
Chemical imaging with HCO$^+$ is in principle a promising method to image the water snowline as it has been used succesfully in a protostellar envelope \citep{vantHoff2018water}. This work examines its application in older protoplanetary disks.
The HCO$^+$ abundance is modelled using two chemical networks. The first one, network NW, does not include reactions with water, in contrast to the second one, network W. Predictions for the radial emission profiles of HCO$^+$ and H$^{13}$CO$^+$ are made to examine the validity of HCO$^+$ as a tracer of the water snowline. 
Moreover, archival observations of V883 Ori are examined and used to put constraints on the water snowline location in V883 Ori and make predictions for future high resolution observations of HCO$^+$ in the disk around this outbursting star. 

Based on our models the following conclusions can be drawn:
\begin{itemize}
\item[$\bullet$] The HCO$^+$ abundance jumps two orders of magnitude around the water snowline. 
\item[$\bullet$] In addition to the water snowline, the optical depth of the continuum emission and molecular excitation effects for the low $J$-levels contribute significantly to the decrease in the H$^{13}$CO$^+$ and HCO$^+$ flux in the inner parts of the disk and result in ring shaped emission. 
Therefore the effect of the continuum optical depth needs to be checked observationally and the effects of the molecular excitation and HCO$^+$ abundance need to be modelled in detail. Outbursting sources are the best targets, as the snowline is shifted to larger radii, where the dust optical depth is lower.
\item[$\bullet$] HCO$^+$ and H$^{13}$CO$^+$ are equally good as a tracers of the water snowline but the main isotopologue HCO$^+$ is more readily observable. 
\item[$\bullet$] For both HCO$^+$ and H$^{13}$CO$^+$, the $J=2-1$ transition is preferred  because it provides the best balance between brightness and effects of the continuum optical depth. Moreover, it is not expected to be blended with emission from complex organic molecules, unlike the $J=3-2$ and $J=4-3$ transitions of H$^{13}$CO$^+$. 
\item[$\bullet$] Our analysis of the observations of the HCO$^+$ $J=3-2$ transition and complex organic molecules suggest that HCO$^+$ is tracing the water snowline in V883 Ori. Based on our models, the snowline is located around 100~AU and is not correlated with the opacity change in the continuum emission observed at 42~AU.
\end{itemize}

Our results thus show that HCO$^+$ and H$^{13}$CO$^+$ can be used to trace the water snowline in warm protoplanetary disks such as those found around luminous stars. Determining the snowline location in these sources is important to understand the process of planet formation and composition.

\begin{acknowledgements}
We thank the referee and the editor for the constructive comments, and Catherine Walsh and Jeong-Eun Lee for useful discussions. 
Astrochemistry in Leiden is supported by the Netherlands Research School for Astronomy (NOVA).
M.L.R.H acknowledges support from a Huygens fellowship from Leiden University, and from the Michigan Society of Fellows. 
L.T. is supported by NWO grant 614.001.352.
This paper makes use of the following ALMA data: ADS/JAO.ALMA\#2017.1.01066.T. and ADS/JAO.ALMA\#2018.1.01131.S. ALMA is a partnership of ESO (representing its member states), NSF (USA) and NINS (Japan), together with NRC (Canada), MOST and ASIAA (Taiwan), and KASI (Republic of Korea), in cooperation with the Republic of Chile. The Joint ALMA Observatory is operated by ESO, AUI/NRAO and NAOJ. 

\end{acknowledgements}

\bibliographystyle{aa}
\bibliography{Refs.bib}

\begin{thebibliography}{106}
\expandafter\ifx\csname natexlab\endcsname\relax\def\natexlab#1{#1}\fi

\bibitem[{{Adams} {et~al.}(1978){Adams}, {Smith}, \& {Grief}}]{Adams1978}
{Adams}, N.~G., {Smith}, D., \& {Grief}, D. 1978, International Journal of Mass
  Spectrometry and Ion Processes, 26, 405

\bibitem[{{Ag{\'u}ndez} {et~al.}(2018){Ag{\'u}ndez}, {Roueff}, {Le Petit}, \&
  {Le Bourlot}}]{Agundez2018}
{Ag{\'u}ndez}, M., {Roueff}, E., {Le Petit}, F., \& {Le Bourlot}, J. 2018,
  \aap, 616, A19

\bibitem[{{Aikawa} {et~al.}(2015){Aikawa}, {Furuya}, {Nomura}, \&
  {Qi}}]{Aikawa2015}
{Aikawa}, Y., {Furuya}, K., {Nomura}, H., \& {Qi}, C. 2015, \apj, 807, 120

\bibitem[{{Andrews}(2020)}]{Andrews2020}
{Andrews}, S.~M. 2020, arXiv e-prints, arXiv:2001.05007

\bibitem[{{Andrews} {et~al.}(2011){Andrews}, {Wilner}, {Espaillat}, {Hughes},
  {Dullemond}, {McClure}, {Qi}, \& {Brown}}]{Andrews2011}
{Andrews}, S.~M., {Wilner}, D.~J., {Espaillat}, C., {et~al.} 2011, \apj, 732,
  42

\bibitem[{{Andrews} {et~al.}(2010){Andrews}, {Wilner}, {Hughes}, {Qi}, \&
  {Dullemond}}]{Andrews2010}
{Andrews}, S.~M., {Wilner}, D.~J., {Hughes}, A.~M., {Qi}, C., \& {Dullemond},
  C.~P. 2010, \apj, 723, 1241

\bibitem[{{Anicich} {et~al.}(1975){Anicich}, {Futrell}, {Huntress}, \&
  {Kim}}]{Anicich1975}
{Anicich}, V.~G., {Futrell}, J.~H., {Huntress}, Wesley~T., J., \& {Kim}, J.~K.
  1975, International Journal of Mass Spectrometry and Ion Processes, 18, 63

\bibitem[{{Banzatti} {et~al.}(2015){Banzatti}, {Pinilla}, {Ricci},
  {Pontoppidan}, {Birnstiel}, \& {Ciesla}}]{Banzatti2015}
{Banzatti}, A., {Pinilla}, P., {Ricci}, L., {et~al.} 2015, \apjl, 815, L15

\bibitem[{{Banzatti} {et~al.}(2017){Banzatti}, {Pontoppidan}, {Salyk},
  {Herczeg}, {van Dishoeck}, \& {Blake}}]{Banzatti2017}
{Banzatti}, A., {Pontoppidan}, K.~M., {Salyk}, C., {et~al.} 2017, \apj, 834,
  152

\bibitem[{{Bergin} {et~al.}(2010){Bergin}, {Hogerheijde}, {Brinch}, {Fogel},
  {Y{\i}ld{\i}z}, {Kristensen}, {van Dishoeck}, {Bell}, {Blake}, {Cernicharo},
  {Dominik}, {Lis}, {Melnick}, {Neufeld}, {Pani{\'c}}, {Pearson}, {Bachiller},
  {Baudry}, {Benedettini}, {Benz}, {Bjerkeli}, {Bontemps}, {Braine},
  {Bruderer}, {Caselli}, {Codella}, {Daniel}, {di Giorgio}, {Doty}, {Encrenaz},
  {Fich}, {Fuente}, {Giannini}, {Goicoechea}, {de Graauw}, {Helmich},
  {Herczeg}, {Herpin}, {Jacq}, {Johnstone}, {J{\o}rgensen}, {Larsson},
  {Liseau}, {Marseille}, {McCoey}, {Nisini}, {Olberg}, {Parise}, {Plume},
  {Risacher}, {Santiago-Garc{\'\i}a}, {Saraceno}, {Shipman}, {Tafalla}, {van
  Kempen}, {Visser}, {Wampfler}, {Wyrowski}, {van der Tak}, {Jellema},
  {Tielens}, {Hartogh}, {St{\"u}tzki}, \& {Szczerba}}]{Bergin2010}
{Bergin}, E.~A., {Hogerheijde}, M.~R., {Brinch}, C., {et~al.} 2010, \aap, 521,
  L33

\bibitem[{{Bergin} {et~al.}(1998){Bergin}, {Melnick}, \&
  {Neufeld}}]{Bergin1998}
{Bergin}, E.~A., {Melnick}, G.~J., \& {Neufeld}, D.~A. 1998, \apj, 499, 777

\bibitem[{{Blevins} {et~al.}(2016){Blevins}, {Pontoppidan}, {Banzatti},
  {Zhang}, {Najita}, {Carr}, {Salyk}, \& {Blake}}]{Blevins2016}
{Blevins}, S.~M., {Pontoppidan}, K.~M., {Banzatti}, A., {et~al.} 2016, \apj,
  818, 22

\bibitem[{{Booth} \& {Ilee}(2020)}]{Booth2020}
{Booth}, A.~S. \& {Ilee}, J.~D. 2020, \mnras, 493, L108

\bibitem[{{Booth} {et~al.}(2019){Booth}, {Walsh}, {Ilee}, {Notsu}, {Qi},
  {Nomura}, \& {Akiyama}}]{Booth2019}
{Booth}, A.~S., {Walsh}, C., {Ilee}, J.~D., {et~al.} 2019, \apjl, 882, L31

\bibitem[{{Bosman} {et~al.}(2018){Bosman}, {Tielens}, \& {van
  Dishoeck}}]{Bosman2018}
{Bosman}, A.~D., {Tielens}, A. G.~G.~M., \& {van Dishoeck}, E.~F. 2018, \aap,
  611, A80

\bibitem[{Botschwina {et~al.}(1993)Botschwina, Horn, Flügge, \&
  Seeger}]{Botschwina1993}
Botschwina, P., Horn, M., Flügge, J., \& Seeger, S. 1993, J. Chem. Soc.{,}
  Faraday Trans., 89, 2219

\bibitem[{{Bruderer}(2013)}]{Bruderer2013}
{Bruderer}, S. 2013, \aap, 559, A46

\bibitem[{{Bruderer} {et~al.}(2009){Bruderer}, {Doty}, \&
  {Benz}}]{Bruderer2009}
{Bruderer}, S., {Doty}, S.~D., \& {Benz}, A.~O. 2009, \apjs, 183, 179

\bibitem[{{Bruderer} {et~al.}(2012){Bruderer}, {van Dishoeck}, {Doty}, \&
  {Herczeg}}]{Bruderer2012}
{Bruderer}, S., {van Dishoeck}, E.~F., {Doty}, S.~D., \& {Herczeg}, G.~J. 2012,
  \aap, 541, A91

\bibitem[{{Bryden} {et~al.}(1999){Bryden}, {Chen}, {Lin}, {Nelson}, \&
  {Papaloizou}}]{Bryden1999}
{Bryden}, G., {Chen}, X., {Lin}, D.~N.~C., {Nelson}, R.~P., \& {Papaloizou},
  J.~C.~B. 1999, \apj, 514, 344

\bibitem[{{Carney} {et~al.}(2018){Carney}, {Fedele}, {Hogerheijde}, {Favre},
  {Walsh}, {Bruderer}, {Miotello}, {Murillo}, {Klaassen}, {Henning}, \& {van
  Dishoeck}}]{Carney2018}
{Carney}, M.~T., {Fedele}, D., {Hogerheijde}, M.~R., {et~al.} 2018, \aap, 614,
  A106

\bibitem[{{Carr} \& {Najita}(2008)}]{Carr2008}
{Carr}, J.~S. \& {Najita}, J.~R. 2008, Science, 319, 1504

\bibitem[{{Caselli} {et~al.}(1998){Caselli}, {Walmsley}, {Terzieva}, \&
  {Herbst}}]{Caselli1998}
{Caselli}, P., {Walmsley}, C.~M., {Terzieva}, R., \& {Herbst}, E. 1998, \apj,
  499, 234

\bibitem[{{Cazzoletti} {et~al.}(2018){Cazzoletti}, {van Dishoeck}, {Visser},
  {Facchini}, \& {Bruderer}}]{Cazzoletti2018}
{Cazzoletti}, P., {van Dishoeck}, E.~F., {Visser}, R., {Facchini}, S., \&
  {Bruderer}, S. 2018, \aap, 609, A93

\bibitem[{{Cieza} {et~al.}(2016){Cieza}, {Casassus}, {Tobin}, {Bos},
  {Williams}, {Perez}, {Zhu}, {Caceres}, {Canovas}, {Dunham}, {Hales},
  {Prieto}, {Principe}, {Schreiber}, {Ruiz-Rodriguez}, \& {Zurlo}}]{Cieza2016}
{Cieza}, L.~A., {Casassus}, S., {Tobin}, J., {et~al.} 2016, \nat, 535, 258

\bibitem[{{D'Alessio} {et~al.}(2006){D'Alessio}, {Calvet}, {Hartmann},
  {Franco-Hern{\'a}ndez}, \& {Serv{\'\i}n}}]{DAlessio2006}
{D'Alessio}, P., {Calvet}, N., {Hartmann}, L., {Franco-Hern{\'a}ndez}, R., \&
  {Serv{\'\i}n}, H. 2006, \apj, 638, 314

\bibitem[{{Dong} {et~al.}(2018){Dong}, {Liu}, {Eisner}, {Andrews}, {Fung},
  {Zhu}, {Chiang}, {Hashimoto}, {Liu}, {Casassus}, {Esposito}, {Hasegawa},
  {Muto}, {Pavlyuchenkov}, {Wilner}, {Akiyama}, {Tamura}, \&
  {Wisniewski}}]{Dong2018}
{Dong}, R., {Liu}, S.-y., {Eisner}, J., {et~al.} 2018, \apj, 860, 124

\bibitem[{{Dr{\k{a}}{\.z}kowska} \& {Alibert}(2017)}]{Drazkowska2017}
{Dr{\k{a}}{\.z}kowska}, J. \& {Alibert}, Y. 2017, \aap, 608, A92

\bibitem[{{Du} {et~al.}(2017){Du}, {Bergin}, {Hogerheijde}, {van Dishoeck},
  {Blake}, {Bruderer}, {Cleeves}, {Dominik}, {Fedele}, {Lis}, {Melnick},
  {Neufeld}, {Pearson}, \& {Y{\i}ld{\i}z}}]{Du2017}
{Du}, F., {Bergin}, E.~A., {Hogerheijde}, M., {et~al.} 2017, \apj, 842, 98

\bibitem[{{Eistrup} {et~al.}(2016){Eistrup}, {Walsh}, \& {van
  Dishoeck}}]{Eistrup2016}
{Eistrup}, C., {Walsh}, C., \& {van Dishoeck}, E.~F. 2016, \aap, 595, A83

\bibitem[{{Eistrup} {et~al.}(2018){Eistrup}, {Walsh}, \& {van
  Dishoeck}}]{Eistrup2018}
{Eistrup}, C., {Walsh}, C., \& {van Dishoeck}, E.~F. 2018, \aap, 613, A14

\bibitem[{{Fedele} {et~al.}(2017){Fedele}, {Carney}, {Hogerheijde}, {Walsh},
  {Miotello}, {Klaassen}, {Bruderer}, {Henning}, \& {van
  Dishoeck}}]{Fedele2017}
{Fedele}, D., {Carney}, M., {Hogerheijde}, M.~R., {et~al.} 2017, \aap, 600, A72

\bibitem[{{Flower}(1999)}]{Flower1999}
{Flower}, D.~R. 1999, \mnras, 305, 651

\bibitem[{{Fraser} {et~al.}(2001){Fraser}, {Collings}, {McCoustra}, \&
  {Williams}}]{Fraser2001}
{Fraser}, H.~J., {Collings}, M.~P., {McCoustra}, M.~R.~S., \& {Williams}, D.~A.
  2001, \mnras, 327, 1165

\bibitem[{{Furlan} {et~al.}(2016){Furlan}, {Fischer}, {Ali}, {Stutz}, {Stanke},
  {Tobin}, {Megeath}, {Osorio}, {Hartmann}, \& {Calvet}}]{Furlan2016}
{Furlan}, E., {Fischer}, W.~J., {Ali}, B., {et~al.} 2016, \apjs, 224, 5

\bibitem[{{Harsono} {et~al.}(2015){Harsono}, {Bruderer}, \& {van
  Dishoeck}}]{Harsono2015}
{Harsono}, D., {Bruderer}, S., \& {van Dishoeck}, E.~F. 2015, \aap, 582, A41

\bibitem[{{Harsono} {et~al.}(2020){Harsono}, {Persson}, {Ramos}, {Murillo},
  {Maud}, {Hogerheijde}, {Bosman}, {Kristensen}, {J{\o}rgensen}, {Bergin},
  {Visser}, {Mottram}, \& {van Dishoeck}}]{Harsono2020}
{Harsono}, D., {Persson}, M.~V., {Ramos}, A., {et~al.} 2020, \aap, 636, A26

\bibitem[{{Hartmann} {et~al.}(1998){Hartmann}, {Calvet}, {Gullbring}, \&
  {D'Alessio}}]{Hartmann1998}
{Hartmann}, L., {Calvet}, N., {Gullbring}, E., \& {D'Alessio}, P. 1998, \apj,
  495, 385

\bibitem[{{Hasegawa} {et~al.}(1992){Hasegawa}, {Herbst}, \&
  {Leung}}]{Hasegawa1992}
{Hasegawa}, T.~I., {Herbst}, E., \& {Leung}, C.~M. 1992, \apjs, 82, 167

\bibitem[{{Hildebrand}(1983)}]{Hildebrand1983}
{Hildebrand}, R.~H. 1983, \qjras, 24, 267

\bibitem[{{Hogerheijde} {et~al.}(2011){Hogerheijde}, {Bergin}, {Brinch},
  {Cleeves}, {Fogel}, {Blake}, {Dominik}, {Lis}, {Melnick}, {Neufeld},
  {Pani{\'c}}, {Pearson}, {Kristensen}, {Y{\i}ld{\i}z}, \& {van
  Dishoeck}}]{Hogerheijde2011}
{Hogerheijde}, M.~R., {Bergin}, E.~A., {Brinch}, C., {et~al.} 2011, Science,
  334, 338

\bibitem[{{Hsieh} {et~al.}(2019){Hsieh}, {Murillo}, {Belloche}, {Hirano},
  {Walsh}, {van Dishoeck}, {J{\o}rgensen}, \& {Lai}}]{Hsieh2019}
{Hsieh}, T.-H., {Murillo}, N.~M., {Belloche}, A., {et~al.} 2019, \apj, 884, 149

\bibitem[{{Huang} {et~al.}(2018){Huang}, {Andrews}, {Dullemond}, {Isella},
  {P{\'e}rez}, {Guzm{\'a}n}, {{\"O}berg}, {Zhu}, {Zhang}, {Bai}, {Benisty},
  {Birnstiel}, {Carpenter}, {Hughes}, {Ricci}, {Weaver}, \&
  {Wilner}}]{Huang2018DSHARPII}
{Huang}, J., {Andrews}, S.~M., {Dullemond}, C.~P., {et~al.} 2018, \apj, 869,
  L42

\bibitem[{Isella {et~al.}(2016)Isella, Guidi, Testi, Liu, Li, Li, Weaver,
  Boehler, Carperter, De~Gregorio-Monsalvo, Manara, Natta, P\'erez, Ricci,
  Sargent, Tazzari, \& Turner}]{Isella2016}
Isella, A., Guidi, G., Testi, L., {et~al.} 2016, Phys. Rev. Lett., 117, 251101

\bibitem[{{J{\o}rgensen} {et~al.}(2013){J{\o}rgensen}, {Visser}, {Sakai},
  {Bergin}, {Brinch}, {Harsono}, {Lindberg}, {van Dishoeck}, {Yamamoto},
  {Bisschop}, \& {Persson}}]{Jorgensen2013}
{J{\o}rgensen}, J.~K., {Visser}, R., {Sakai}, N., {et~al.} 2013, \apj, 779, L22

\bibitem[{{Kama} {et~al.}(2016){Kama}, {Bruderer}, {van Dishoeck},
  {Hogerheijde}, {Folsom}, {Miotello}, {Fedele}, {Belloche}, {G{\"u}sten}, \&
  {Wyrowski}}]{Kama2016}
{Kama}, M., {Bruderer}, S., {van Dishoeck}, E.~F., {et~al.} 2016, \aap, 592,
  A83

\bibitem[{{Kama} {et~al.}(2020){Kama}, {Trapman}, {Fedele}, {Bruderer},
  {Hogerheijde}, {Miotello}, {van Dishoeck}, {Clarke}, \& {Bergin}}]{Kama2020}
{Kama}, M., {Trapman}, L., {Fedele}, D., {et~al.} 2020, \aap, 634, A88

\bibitem[{{Kim} {et~al.}(1974){Kim}, {Theard}, \& {Huntress}}]{Kim1974}
{Kim}, J.~K., {Theard}, L.~P., \& {Huntress}, W.~T., J. 1974, International
  Journal of Mass Spectrometry and Ion Processes, 15, 223

\bibitem[{Kimura {et~al.}(2020)Kimura, Wada, Kobayashi, Senshu, Hirai, Yoshida,
  Kobayashi, Hong, Arai, Ishibashi, \& Yamada}]{Kimura2020}
Kimura, H., Wada, K., Kobayashi, H., {et~al.} 2020, Monthly Notices of the
  Royal Astronomical Society, staa2467

\bibitem[{{Klippenstein} {et~al.}(2010){Klippenstein}, {Georgievskii}, \&
  {McCall}}]{Klippenstein2010}
{Klippenstein}, S.~J., {Georgievskii}, Y., \& {McCall}, B.~J. 2010, Journal of
  Physical Chemistry A, 114, 278

\bibitem[{{Kounkel} {et~al.}(2017){Kounkel}, {Hartmann}, {Loinard},
  {Ortiz-Le{\'o}n}, {Mioduszewski}, {Rodr{\'\i}guez}, {Dzib}, {Torres}, {Pech},
  {Galli}, {Rivera}, {Boden}, {Evans}, {Brice{\~n}o}, \& {Tobin}}]{Kounkel2017}
{Kounkel}, M., {Hartmann}, L., {Loinard}, L., {et~al.} 2017, \apj, 834, 142

\bibitem[{{Lee} {et~al.}(2019){Lee}, {Lee}, {Baek}, {Aikawa}, {Cieza}, {Yoon},
  {Herczeg}, {Johnstone}, \& {Casassus}}]{Lee2019}
{Lee}, J.-E., {Lee}, S., {Baek}, G., {et~al.} 2019, Nature Astronomy, 3, 314

\bibitem[{{Lepp} {et~al.}(1987){Lepp}, {Dalgarno}, \& {Sternberg}}]{Lepp1987}
{Lepp}, S., {Dalgarno}, A., \& {Sternberg}, A. 1987, \apj, 321, 383

\bibitem[{{Long} {et~al.}(2018){Long}, {Pinilla}, {Herczeg}, {Harsono},
  {Dipierro}, {Pascucci}, {Hendler}, {Tazzari}, {Ragusa}, {Salyk}, {Edwards},
  {Lodato}, {van de Plas}, {Johnstone}, {Liu}, {Boehler}, {Cabrit}, {Manara},
  {Menard}, {Mulders}, {Nisini}, {Fischer}, {Rigliaco}, {Banzatti}, {Avenhaus},
  \& {Gully-Santiago}}]{Long2018}
{Long}, F., {Pinilla}, P., {Herczeg}, G.~J., {et~al.} 2018, \apj, 869, 17

\bibitem[{{Lynden-Bell} \& {Pringle}(1974)}]{LyndenBell1974}
{Lynden-Bell}, D. \& {Pringle}, J.~E. 1974, \mnras, 168, 603

\bibitem[{{Mandell} {et~al.}(2012){Mandell}, {Bast}, {van Dishoeck}, {Blake},
  {Salyk}, {Mumma}, \& {Villanueva}}]{Mandell2012}
{Mandell}, A.~M., {Bast}, J., {van Dishoeck}, E.~F., {et~al.} 2012, \apj, 747,
  92

\bibitem[{{Mathews} {et~al.}(2013){Mathews}, {Klaassen}, {Juh{\'a}sz},
  {Harsono}, {Chapillon}, {van Dishoeck}, {Espada}, {de Gregorio-Monsalvo},
  {Hales}, \& {Hogerheijde}}]{Mathews2013}
{Mathews}, G.~S., {Klaassen}, P.~D., {Juh{\'a}sz}, A., {et~al.} 2013, \aap,
  557, A132

\bibitem[{{Mathis} {et~al.}(1977){Mathis}, {Rumpl}, \&
  {Nordsieck}}]{Mathis1977}
{Mathis}, J.~S., {Rumpl}, W., \& {Nordsieck}, K.~H. 1977, \apj, 217, 425

\bibitem[{{McCall} {et~al.}(2004){McCall}, {Huneycutt}, {Saykally}, {Djuric},
  {Dunn}, {Semaniak}, {Novotny}, {Al-Khalili}, {Ehlerding}, {Hellberg},
  {Kalhori}, {Neau}, {Thomas}, {Paal}, {{\"O}sterdahl}, \&
  {Larsson}}]{McCall2004}
{McCall}, B.~J., {Huneycutt}, A.~J., {Saykally}, R.~J., {et~al.} 2004, \pra,
  70, 052716

\bibitem[{{McElroy} {et~al.}(2013){McElroy}, {Walsh}, {Markwick}, {Cordiner},
  {Smith}, \& {Millar}}]{McElroy2013}
{McElroy}, D., {Walsh}, C., {Markwick}, A.~J., {et~al.} 2013, \aap, 550, A36

\bibitem[{{Milam} {et~al.}(2005){Milam}, {Savage}, {Brewster}, {Ziurys}, \&
  {Wyckoff}}]{Milam2005}
{Milam}, S.~N., {Savage}, C., {Brewster}, M.~A., {Ziurys}, L.~M., \& {Wyckoff},
  S. 2005, \apj, 634, 1126

\bibitem[{{Mitchell}(1990)}]{Mitchell1990}
{Mitchell}, J.~B.~A. 1990, \physrep, 186, 215

\bibitem[{{Notsu} {et~al.}(2019){Notsu}, {Akiyama}, {Booth}, {Nomura}, {Walsh},
  {Hirota}, {Honda}, {Tsukagoshi}, \& {Millar}}]{Notsu2019}
{Notsu}, S., {Akiyama}, E., {Booth}, A., {et~al.} 2019, \apj, 875, 96

\bibitem[{{Novotn{\'y}} {et~al.}(2010){Novotn{\'y}}, {Buhr}, {St{\"u}tzel},
  {Mendes}, {Berg}, {Bing}, {Froese}, {Grieser}, {Heber}, {Jordon-Thaden},
  {Krantz}, {Lange}, {Lestinsky}, {Novotny}, {Menk}, {Orlov}, {Petrignani},
  {Rappaport}, {Shornikov}, {Schwalm}, {Zajfman}, \& {Wolf}}]{Novotny2010}
{Novotn{\'y}}, O., {Buhr}, H., {St{\"u}tzel}, J., {et~al.} 2010, Journal of
  Physical Chemistry A, 114, 4870

\bibitem[{{{\"O}berg} {et~al.}(2011){{\"O}berg}, {Murray-Clay}, \&
  {Bergin}}]{Oberg2011}
{{\"O}berg}, K.~I., {Murray-Clay}, R., \& {Bergin}, E.~A. 2011, \apj, 743, L16

\bibitem[{{Okuzumi} {et~al.}(2016){Okuzumi}, {Momose}, {Sirono}, {Kobayashi},
  \& {Tanaka}}]{Okuzumi2016}
{Okuzumi}, S., {Momose}, M., {Sirono}, S.-i., {Kobayashi}, H., \& {Tanaka}, H.
  2016, \apj, 821, 82

\bibitem[{{Phillips} {et~al.}(1992){Phillips}, {van Dishoeck}, \&
  {Keene}}]{Phillips1992}
{Phillips}, T.~G., {van Dishoeck}, E.~F., \& {Keene}, J. 1992, \apj, 399, 533

\bibitem[{{Pickering}(1890)}]{Pickering1890}
{Pickering}, E.~C. 1890, Annals of Harvard College Observatory, 18, 1

\bibitem[{{Pickett} {et~al.}(1998){Pickett}, {Poynter}, {Cohen}, {Delitsky},
  {Pearson}, \& {M{\"u}ller}}]{Pickett1998}
{Pickett}, H.~M., {Poynter}, R.~L., {Cohen}, E.~A., {et~al.} 1998, \jqsrt, 60,
  883

\bibitem[{{Pinilla} {et~al.}(2017){Pinilla}, {Pohl}, {Stammler}, \&
  {Birnstiel}}]{Pinilla2017}
{Pinilla}, P., {Pohl}, A., {Stammler}, S.~M., \& {Birnstiel}, T. 2017, \apj,
  845, 68

\bibitem[{{Qi} {et~al.}(2015){Qi}, {{\"O}berg}, {Andrews}, {Wilner}, {Bergin},
  {Hughes}, {Hogherheijde}, \& {D'Alessio}}]{Qi2015}
{Qi}, C., {{\"O}berg}, K.~I., {Andrews}, S.~M., {et~al.} 2015, \apj, 813, 128

\bibitem[{{Qi} {et~al.}(2019){Qi}, {{\"O}berg}, {Espaillat}, {Robinson},
  {Andrews}, {Wilner}, {Blake}, {Bergin}, \& {Cleeves}}]{Qi2019}
{Qi}, C., {{\"O}berg}, K.~I., {Espaillat}, C.~C., {et~al.} 2019, \apj, 882, 160

\bibitem[{{Qi} {et~al.}(2013){Qi}, {{\"O}berg}, {Wilner}, {D'Alessio},
  {Bergin}, {Andrews}, {Blake}, {Hogerheijde}, \& {van Dishoeck}}]{Qi2013}
{Qi}, C., {{\"O}berg}, K.~I., {Wilner}, D.~J., {et~al.} 2013, Science, 341, 630

\bibitem[{{Ru{\'\i}z-Rodr{\'\i}guez} {et~al.}(2017){Ru{\'\i}z-Rodr{\'\i}guez},
  {Cieza}, {Williams}, {Principe}, {Tobin}, {Zhu}, \&
  {Zurlo}}]{RuizRodriguez2017}
{Ru{\'\i}z-Rodr{\'\i}guez}, D., {Cieza}, L.~A., {Williams}, J.~P., {et~al.}
  2017, \mnras, 468, 3266

\bibitem[{{Salinas} {et~al.}(2016){Salinas}, {Hogerheijde}, {Bergin},
  {Cleeves}, {Brinch}, {Blake}, {Lis}, {Melnick}, {Pani{\'c}}, {Pearson},
  {Kristensen}, {Y{\i}ld{\i}z}, \& {van Dishoeck}}]{Salinas2016}
{Salinas}, V.~N., {Hogerheijde}, M.~R., {Bergin}, E.~A., {et~al.} 2016, \aap,
  591, A122

\bibitem[{{Salyk} {et~al.}(2019){Salyk}, {Lacy}, {Richter}, {Zhang},
  {Pontoppidan}, {Carr}, {Najita}, \& {Blake}}]{Salyk2019}
{Salyk}, C., {Lacy}, J., {Richter}, M., {et~al.} 2019, \apj, 874, 24

\bibitem[{{Salyk} {et~al.}(2008){Salyk}, {Pontoppidan}, {Blake}, {Lahuis}, {van
  Dishoeck}, \& {Evans}}]{Salyk2008}
{Salyk}, C., {Pontoppidan}, K.~M., {Blake}, G.~A., {et~al.} 2008, \apjl, 676,
  L49

\bibitem[{{Sandell} \& {Weintraub}(2001)}]{Sandell2001}
{Sandell}, G. \& {Weintraub}, D.~A. 2001, \apjs, 134, 115

\bibitem[{{Sch{\"o}ier} {et~al.}(2005){Sch{\"o}ier}, {van der Tak}, {van
  Dishoeck}, \& {Black}}]{Schoier2005}
{Sch{\"o}ier}, F.~L., {van der Tak}, F.~F.~S., {van Dishoeck}, E.~F., \&
  {Black}, J.~H. 2005, \aap, 432, 369

\bibitem[{{Schoonenberg} \& {Ormel}(2017)}]{Schoonenberg2017}
{Schoonenberg}, D. \& {Ormel}, C.~W. 2017, \aap, 602, A21

\bibitem[{{Stevenson} \& {Lunine}(1988)}]{Stevenson1988}
{Stevenson}, D.~J. \& {Lunine}, J.~I. 1988, \icarus, 75, 146

\bibitem[{{Strom} \& {Strom}(1993)}]{Strom1993}
{Strom}, K.~M. \& {Strom}, S.~E. 1993, \apj, 412, L63

\bibitem[{{Teague} {et~al.}(2016){Teague}, {Guilloteau}, {Semenov}, {Henning},
  {Dutrey}, {Pi{\'e}tu}, {Birnstiel}, {Chapillon}, {Hollenbach}, \&
  {Gorti}}]{Teague2016}
{Teague}, R., {Guilloteau}, S., {Semenov}, D., {et~al.} 2016, \aap, 592, A49

\bibitem[{{Teague} {et~al.}(2015){Teague}, {Semenov}, {Guilloteau}, {Henning},
  {Dutrey}, {Wakelam}, {Chapillon}, \& {Pietu}}]{Teague2015}
{Teague}, R., {Semenov}, D., {Guilloteau}, S., {et~al.} 2015, \aap, 574, A137

\bibitem[{{Theard} \& {Huntress}(1974)}]{Theard1974}
{Theard}, L.~P. \& {Huntress}, W.~T. 1974, \jcp, 60, 2840

\bibitem[{{Tobin} {et~al.}(2020){Tobin}, {Sheehan}, {Megeath},
  {D{\'\i}az-Rodr{\'\i}guez}, {Offner}, {Murillo}, {van 't Hoff}, {van
  Dishoeck}, {Osorio}, {Anglada}, {Furlan}, {Stutz}, {Reynolds}, {Karnath},
  {Fischer}, {Persson}, {Looney}, {Li}, {Stephens}, {Chand ler}, {Cox},
  {Dunham}, {Tychoniec}, {Kama}, {Kratter}, {Kounkel}, {Mazur}, {Maud},
  {Patel}, {Perez}, {Sadavoy}, {Segura-Cox}, {Sharma}, {Stephenson}, {Watson},
  \& {Wyrowski}}]{Tobin2020}
{Tobin}, J.~J., {Sheehan}, P.~D., {Megeath}, S.~T., {et~al.} 2020, \apj, 890,
  130

\bibitem[{{Tychoniec} {et~al.}(2020){Tychoniec}, {Manara}, {Rosotti}, {van
  Dishoeck}, {Cridland }, {Hsieh}, {Murillo}, {Segura-Cox}, {van Terwisga}, \&
  {Tobin}}]{Tychoniec2020}
{Tychoniec}, {\L}., {Manara}, C.~F., {Rosotti}, G.~P., {et~al.} 2020, A\&A,
  640, A19

\bibitem[{{van der Marel} {et~al.}(2015){van der Marel}, {van Dishoeck},
  {Bruderer}, {P{\'e}rez}, \& {Isella}}]{vanderMarel2015}
{van der Marel}, N., {van Dishoeck}, E.~F., {Bruderer}, S., {P{\'e}rez}, L., \&
  {Isella}, A. 2015, \aap, 579, A106

\bibitem[{{van Gelder} {et~al.}(2020){van Gelder}, {Tabone}, {Tychoniec}, {van
  Dishoeck}, {Beuther}, {Boogert}, {Caratti o Garatti}, {Klaassen}, {Linnartz},
  {M{\"u}ller}, \& {Taquet}}]{vanGelder2020}
{van Gelder}, M.~L., {Tabone}, B., {Tychoniec}, {\L}., {et~al.} 2020, \aap,
  639, A87

\bibitem[{{van 't Hoff} {et~al.}(2018{\natexlab{a}}){van 't Hoff}, {Persson},
  {Harsono}, {Taquet}, {J{\o}rgensen}, {Visser}, {Bergin}, \& {van
  Dishoeck}}]{vantHoff2018water}
{van 't Hoff}, M. L.~R., {Persson}, M.~V., {Harsono}, D., {et~al.}
  2018{\natexlab{a}}, \aap, 613, A29

\bibitem[{{van 't Hoff} {et~al.}(2018{\natexlab{b}}){van 't Hoff}, {Tobin},
  {Trapman}, {Harsono}, {Sheehan}, {Fischer}, {Megeath}, \& {van
  Dishoeck}}]{vantHoff2018methanol}
{van 't Hoff}, M. L.~R., {Tobin}, J.~J., {Trapman}, L., {et~al.}
  2018{\natexlab{b}}, \apj, 864, L23

\bibitem[{{van 't Hoff} {et~al.}(2017){van 't Hoff}, {Walsh}, {Kama},
  {Facchini}, \& {van Dishoeck}}]{vantHoff2017}
{van 't Hoff}, M.~L.~R., {Walsh}, C., {Kama}, M., {Facchini}, S., \& {van
  Dishoeck}, E.~F. 2017, \aap, 599, A101

\bibitem[{{van Terwisga} {et~al.}(2018){van Terwisga}, {van Dishoeck},
  {Ansdell}, {van der Marel}, {Testi}, {Williams}, {Facchini}, {Tazzari},
  {Hogerheijde}, {Trapman}, {Manara}, {Miotello}, {Maud}, \&
  {Harsono}}]{vanTerwisga2018}
{van Terwisga}, S.~E., {van Dishoeck}, E.~F., {Ansdell}, M., {et~al.} 2018,
  \aap, 616, A88

\bibitem[{{van Zadelhoff} {et~al.}(2003){van Zadelhoff}, {Aikawa},
  {Hogerheijde}, \& {van Dishoeck}}]{vanZadelhoff2003}
{van Zadelhoff}, G.~J., {Aikawa}, Y., {Hogerheijde}, M.~R., \& {van Dishoeck},
  E.~F. 2003, \aap, 397, 789

\bibitem[{{Visser} \& {Bergin}(2012)}]{Visser2012}
{Visser}, R. \& {Bergin}, E.~A. 2012, \apjl, 754, L18

\bibitem[{{Visser} {et~al.}(2015){Visser}, {Bergin}, \&
  {J{\o}rgensen}}]{Visser2015}
{Visser}, R., {Bergin}, E.~A., \& {J{\o}rgensen}, J.~K. 2015, \aap, 577, A102

\bibitem[{{Vorobyov} {et~al.}(2013){Vorobyov}, {Baraffe}, {Harries}, \&
  {Chabrier}}]{Vorobyov2013}
{Vorobyov}, E.~I., {Baraffe}, I., {Harries}, T., \& {Chabrier}, G. 2013, \aap,
  557, A35

\bibitem[{{Walsh} {et~al.}(2013){Walsh}, {Millar}, \& {Nomura}}]{Walsh2013}
{Walsh}, C., {Millar}, T.~J., \& {Nomura}, H. 2013, \apjl, 766, L23

\bibitem[{{Walsh} {et~al.}(2012){Walsh}, {Nomura}, {Millar}, \&
  {Aikawa}}]{Walsh2012}
{Walsh}, C., {Nomura}, H., {Millar}, T.~J., \& {Aikawa}, Y. 2012, \apj, 747,
  114

\bibitem[{{Williams} {et~al.}(1998){Williams}, {Bergin}, {Caselli}, {Myers}, \&
  {Plume}}]{Williams1998}
{Williams}, J.~P., {Bergin}, E.~A., {Caselli}, P., {Myers}, P.~C., \& {Plume},
  R. 1998, \apj, 503, 689

\bibitem[{{Woitke} {et~al.}(2016){Woitke}, {Min}, {Pinte}, {Thi}, {Kamp},
  {Rab}, {Anthonioz}, {Antonellini}, {Baldovin-Saavedra}, {Carmona}, {Dominik},
  {Dionatos}, {Greaves}, {G{\"u}del}, {Ilee}, {Liebhart}, {M{\'e}nard},
  {Rigon}, {Waters}, {Aresu}, {Meijerink}, \& {Spaans}}]{Woitke2016}
{Woitke}, P., {Min}, M., {Pinte}, C., {et~al.} 2016, \aap, 586, A103

\bibitem[{{Yen} {et~al.}(2016){Yen}, {Koch}, {Liu}, {Puspitaningrum}, {Hirano},
  {Lee}, \& {Takakuwa}}]{Yen2016}
{Yen}, H.-W., {Koch}, P.~M., {Liu}, H.~B., {et~al.} 2016, \apj, 832, 204

\bibitem[{{Yen} {et~al.}(2018){Yen}, {Koch}, {Manara}, {Miotello}, \&
  {Testi}}]{Yen2018}
{Yen}, H.-W., {Koch}, P.~M., {Manara}, C.~F., {Miotello}, A., \& {Testi}, L.
  2018, \aap, 616, A100

\bibitem[{{Zhang} {et~al.}(2015){Zhang}, {Blake}, \& {Bergin}}]{Zhang2015}
{Zhang}, K., {Blake}, G.~A., \& {Bergin}, E.~A. 2015, \apjl, 806, L7

\bibitem[{{Zhang} {et~al.}(2013){Zhang}, {Pontoppidan}, {Salyk}, \&
  {Blake}}]{Zhang2013}
{Zhang}, K., {Pontoppidan}, K.~M., {Salyk}, C., \& {Blake}, G.~A. 2013, \apj,
  766, 82

\bibitem[{{Zhu} {et~al.}(2014){Zhu}, {Stone}, {Rafikov}, \& {Bai}}]{Zhu2014}
{Zhu}, Z., {Stone}, J.~M., {Rafikov}, R.~R., \& {Bai}, X.-n. 2014, \apj, 785,
  122

\end{thebibliography}

\begin{appendix} 

\section{Freeze-out and desorption coefficients} \label{sec:fdwater}

The reaction coefficients of the freeze-out and desorption of water depend on the local conditions in the disk. The rate coefficient of the freeze-out of water, $k_{\mathrm{f}}$ can be expressed as: 
\begin{align}
k_{\mathrm{f}} &= \langle v \rangle \pi a_{\mathrm{grain}}^2 n(\mathrm{grain}) S \\
& = \sqrt{\frac{k_{\mathrm{B}}T_{\mathrm{gas}}}{m}} \pi a_{\mathrm{grain}}^2 n(\mathrm{grain}) S,
\end{align}
where $\langle v \rangle = \sqrt{k_{\mathrm{B}}T_{\mathrm{gas}}/m}$ is the thermal velocity of gas-phase water molecules with mass $m$ in a gas at temperature $T_{\mathrm{gas}}$, and
$k_{\mathrm{B}}$ is the Boltzmann constant. The \texttt{DALI} models use a distribution of grain sizes, but for the chemistry, only the surface area of the grains matters. Therefore a grain size of $a_{\mathrm{grain}} = 0.1~\mu$m and a number density, $n(\mathrm{grain}) = 10^{-12}\times n(\mathrm{H_2})$ with respect to molecule hydrogen is used. Finally a sticking coefficient, $S$, of 1 is assumed.

The reverse process, the desorption of water ice, is described by the desorption rate coefficient:
\begin{align}
k_{\mathrm{d}} &= \nu_0 e^{-E_{\mathrm{b}}/k_{\mathrm{B}}T_{\mathrm{dust}}} \\
&= \sqrt{\frac{2n_{\mathrm{s}}E_{\mathrm{b}}}{\pi^2 m}} e^{-E_{\mathrm{b}}/k_{\mathrm{B}}T_{\mathrm{dust}}},
\end{align}
with $\nu_0 = \sqrt{2n_{\mathrm{s}}E_{\mathrm{b}}/\pi^2 m}$ the characteristic vibrational frequency of water ice on a grain. Here $n_{\mathrm{s}} = 1.5\times10^{15}$~cm$^{-2}$ is the number density of surface sites where water can bind \citep{Hasegawa1992}. The binding energy, $E_{\mathrm{b}}$, assumed in this work is 5775~K, corresponding to an amorphous water ice substrate \citep{Fraser2001}.

\section{Chemical network}

\subsection{Analytical approximation for network NW} \label{sec:appxHCOp}
Chemical network NW can be used to derive an analytical expression for the HCO$^+$ abundance because of its simplicity. The time evolution of the HCO$^+$ number density is given by the sum of the formation and destruction rates:
\begin{align}
	\frac{d n(\mathrm{HCO^+})}{d t} = k_{\mathrm{4}} n(\mathrm{CO})n(\mathrm{H_3^+}) - k_{\mathrm{e^-}}n(\mathrm{HCO^+})n(\mathrm{e^-}),
\end{align}
with $k_{\mathrm{4}}$ and $k_{\mathrm{e^-}}$ the reaction rates as listed in Table~\ref{tab:reactions}. This can be rewritten to 
\begin{align}
	n(\mathrm{HCO^+}) = \sqrt{\frac{k_{\mathrm{4}}}{k_{\mathrm{e^-}}} n(\mathrm{CO})n(\mathrm{H_3^+})} \label{eq:HCOp_ana1}
\end{align}
under the assumption of steady state. Furthermore, it is assumed that metallic ions can be neglected as electron donors and that HCO$^+$ is the main electron donor. HCO$^+$ has been found to be the dominant molecular ion protoplanetary disks \citep{Teague2015} and the main charge carrier in starless cores (e.g. \citealt{Caselli1998, Williams1998}). 

Following the approach of \citet{Lepp1987}, the number density of H$_3^+$, which is mainly governed by the ionization, can be found in a similar way. The three reactions that regulate the ionization in the disk: ionization by cosmic rays (reaction $\zeta_{\mathrm{c.r.}}$), the ion-molecule reaction to form H$_3^+$ (reaction $k_{\mathrm{2}}$) and dissociative recombination of H$_3^+$ (reaction $k_{\mathrm{3}}$), see also Table.~\ref{tab:reactions}. The first two reactions can be approximated as:
\begin{align}
	\mathrm{H_{2}} + \mathrm{c.r.} \to \mathrm{H_3^+},
\end{align}
as the formation of H$_3^+$ is limited by the cosmic ray ionization rate. Assuming CO is the main destroyer of H$_3^+$ and thus neglecting the dissociative recombination of H$_3^+$, the time evolution and steady state abundance of H$_3^+$ can be expressed as:
\begin{align}
	\frac{d n(\mathrm{H_3^+})}{d t} &= \zeta_{\mathrm{c.r.}} n(\mathrm{H_2}) - k_{\mathrm{4}}n(\mathrm{CO})n(\mathrm{H_3^+})\ \ \ \text{(time-dependent), and} \\
	n(\mathrm{H_3^+}) &= \frac{\zeta_{\mathrm{c.r.}}n(\mathrm{H_2})}{k_{\mathrm{4}}n(\mathrm{CO})}\ \ \ \mathrm{(steady\ state).} \label{eq:H3pana}
\end{align}

Combining Eq.~\ref{eq:HCOp_ana1} and Eq.~\ref{eq:H3pana} gives the analytical approximation of the HCO$^+$ abundance in the disk:
\begin{align}
x(\mathrm{HCO^+}) = \sqrt{\frac{\zeta_{\mathrm{c.r.}}}{k_{\mathrm{e^-}}n(\mathrm{H_2})}}.
\end{align}
The comparison of chemical network NW and W in Section~\ref{sec:resultschem} and Fig.~\ref{fig:HCOpnetworkAD} shows that the HCO$^+$ abundance in network NW is very similar to the HCO$^+$ abundance outside the water snow surface in network W. Therefore, the derived expression for the HCO$^+$ abundance can also be used in this disk region in chemical network W.

\subsection{Initial conditions} \label{sec:appinit}

\begin{figure*}
   \centering
   \includegraphics[width=1\linewidth]{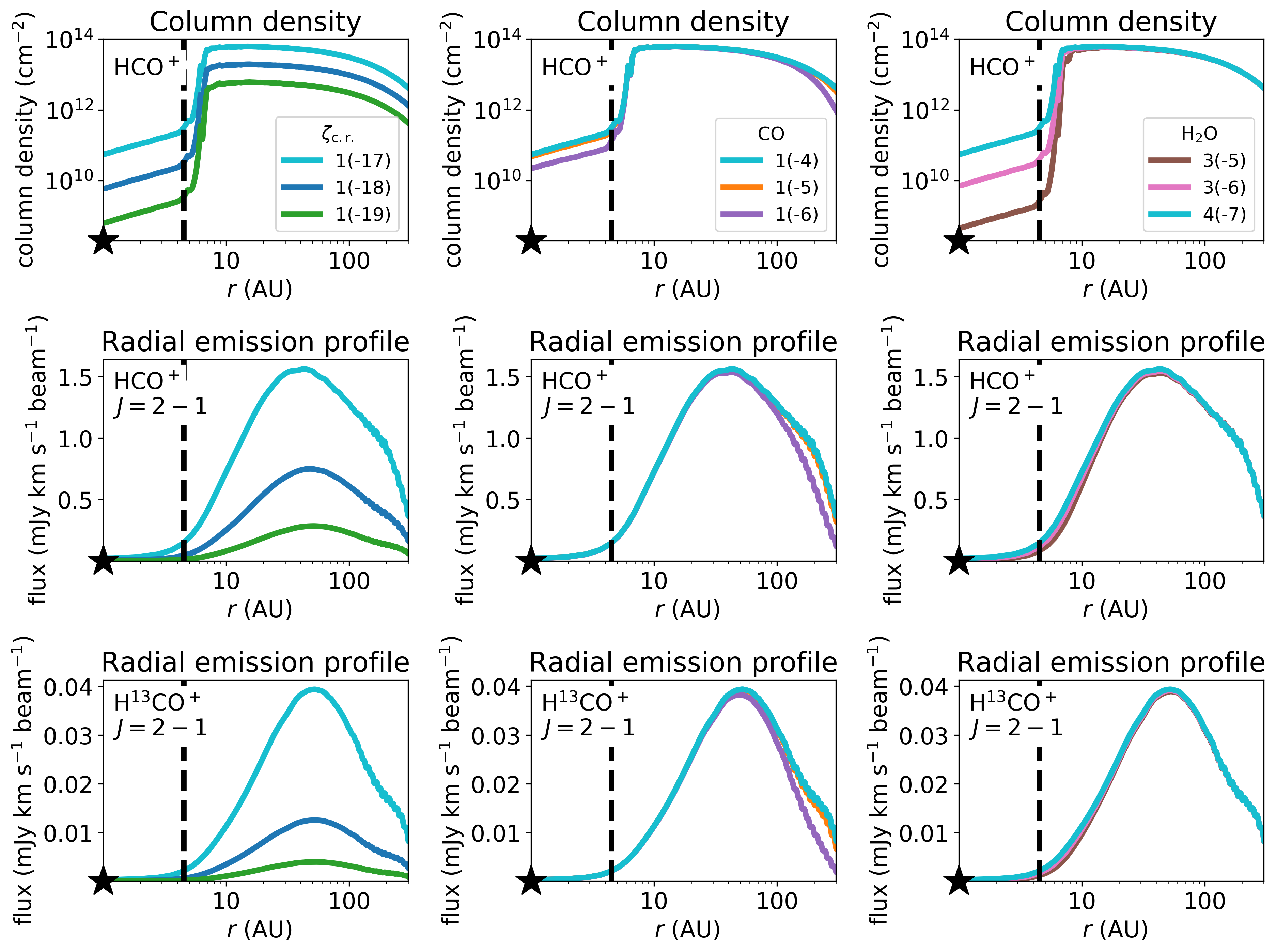}
      \caption{HCO$^+$ column densities for different initial conditions in chemical network W (top row) and the corresponding radial emission profiles for the $J=2-1$ transition of HCO$^+$ (middle row) and H$^{13}$CO$^+$ (bottom row). The left-hand column shows the results for a cosmic ray ionization rate of $10^{-17} \mathrm{s}^{-1}$, $10^{-18} \mathrm{s}^{-1}$ and $10^{-19} \mathrm{s}^{-1}$. The middle column shows the corresponding models for an initial CO abundance of $10^{-4}$, $10^{-5}$ and $10^{-6}$. The right-hand column shows the corresponding models for an initial abundance of gas-phase water of $3\times 10^{-5}$, $3\times 10^{-6}$ and $3.8\times 10^{-7}$. The fiducial model is indicated with the light blue line in each panel and uses an initial abundance of $3.8\times 10^{-7}$ for gas-phase water, $10^{-4}$ for gas-phase CO and $10^{-17} \mathrm{s}^{-1}$ for the cosmic ray ionisation rate. The water snowline is indicated with a dashed black line and the position of the star is indicated by the symbol of a black star. The radial emission profiles are convolved with a $0 \farcs 05$ beam. }
         \label{fig:icchemnet}
\end{figure*}

The effects of the initial conditions on the abundance predicted by chemical network W were discussed in Section~\ref{sec:initial_cond_abu} and expected to be of little importance for HCO$^+$'s ability to trace the water snowline. 
Here, the effects on the corresponding radial emission profiles of the $J=2-1$ transition of HCO$^+$ and H$^{13}$CO$^+$ are discussed and shown in Fig.~\ref{fig:icchemnet}. 

Previously, it was derived that the column density of HCO$^+$ scales with the square root of the cosmic ray ionization rate (see Eq.~\ref{eq:xHCOp}). Similarly, the H$^{13}$CO$^+$ emission scales with the square root of the cosmic ray ionization rate because it is optically thin, see Fig.~\ref{fig:icchemnet}. On the other hand, the HCO$^+$ emission does not scale with the cosmic ray ionization rate as it is optically thick. As the column density of HCO$^+$ decreases, the HCO$^+$ emission becomes less optically thick and approaches the scaling for the column density. 

In Section~\ref{sec:initial_cond_abu}, it was found that the HCO$^+$ abundance or column density does not depend strongly on the initial abundance of CO or H$_2$O. This is is also seen in the radial emission profiles both for HCO$^+$ and H$^{13}$CO$^+$, because the drop in the HCO$^+$ emission in the center is dominated by the effect of the optical depth of the continuum emission and molecular excitation. A small dependence of the HCO$^+$ emission on the initial abundance of gas-phase water is seen, but the abundance of gas-phase water seems to be low in the outer regions of protoplanetary disks \citep{Bergin2010, Du2017, Notsu2019, Harsono2020}. Moreover, the abundance of gas-phase water in the inner disk can be as high as $10^{-2}$ \citep{Bosman2018}.

\begin{table}
\centering
\begin{threeparttable}
	\caption{Initial conditions for chemical network NW (no water) and W (water).}    
    \begin{tabularx}{\linewidth}{p{0.35\columnwidth}p{0.21\columnwidth}p{0.21\columnwidth}p{0.23\columnwidth}}
    \hline
        Model parameter & Initial value NW & Initial value W & Ref. \\ \hline
        $\zeta_{\mathrm{c.r.}}$ & $1(-17$) & $1(-17)$ & \\  
        $T_{\mathrm{gas}}$ & \texttt{DALI} model & \texttt{DALI} model & \\
        $T_{\mathrm{dust}}$ & \texttt{DALI} model & \texttt{DALI} model & \\
        $n(\mathrm{H_2})$ & \texttt{DALI} model & \texttt{DALI} model &\\
        $x(\mathrm{H_2^+)}$ & $0$ & $0$ &\\
        $x(\mathrm{H_3^+})$ & $0$ & $0$ &\\
        $x(\mathrm{H})$ & $0$ & $0$ &\\     
        $x(\mathrm{e^-})$ & $0$ & $0$ &\\ 
        $x(\mathrm{CO})$ & $1(-4)$ & $1(-4)$ &\\ 
        $x(\mathrm{HCO^+})$ & $0$ & $0$ &\\ 
        $x(\mathrm{H_3O^+})$ & $0$ & $0$ &\\ 
        $x(\mathrm{H_2O})$ & $0$ & $3.8(-7)$ & (a) \\
        $x(\mathrm{H_2O (ice)})$ & $0$ & $0$ &\\  \hline
    \end{tabularx}
    \begin{tablenotes}
      \small
      \item \textbf{Notes.} $a(b)$ represents $a\times 10^b$. Abundances are defined with respect to molecular hydrogen. References: (a) \citealt{McElroy2013}.
    \end{tablenotes}
    \label{tab:params_chem_net}
    \end{threeparttable}
\end{table}

\section{HCO$^+$ and H$^{13}$CO$^+$ $J=1-0$, $J=3-2$ and $J=4-3$ transitions} \label{app:Radialcutsotherlines}

\begin{figure*}
   \centering
   \includegraphics[width=1\linewidth]{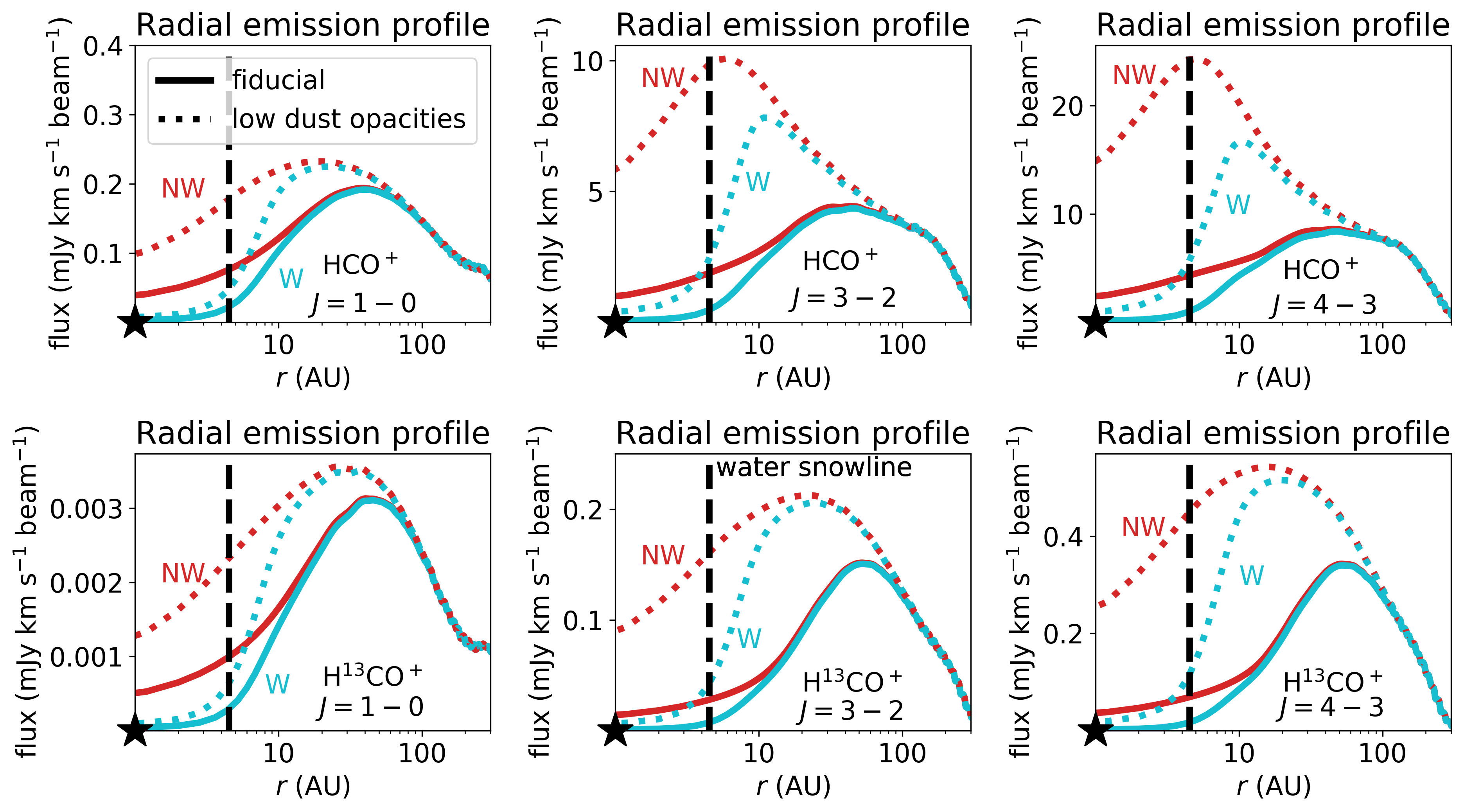}
      \caption{Same as the middle and bottom panel of Fig.~\ref{fig:radialmodelandobs}, but then for the $J=1-0$ (left column), $J=3-2$ (middle column) and $J=4-3$ (right column) transition of HCO$^+$ (top) and H$^{13}$CO$^+$ (bottom). }
         \label{fig:herbigdiffradcuts}
\end{figure*}

Radial emission profiles for the $J=1-0$, $J=3-2$ and $J=4-3$ transitions of HCO$^+$ and H$^{13}$CO$^+$ are presented in Fig.~\ref{fig:herbigdiffradcuts}.

\section{V883 Ori} \label{app:V883Ori_mom0s}

\begin{figure*}
   \centering
   \includegraphics[width=1\linewidth]{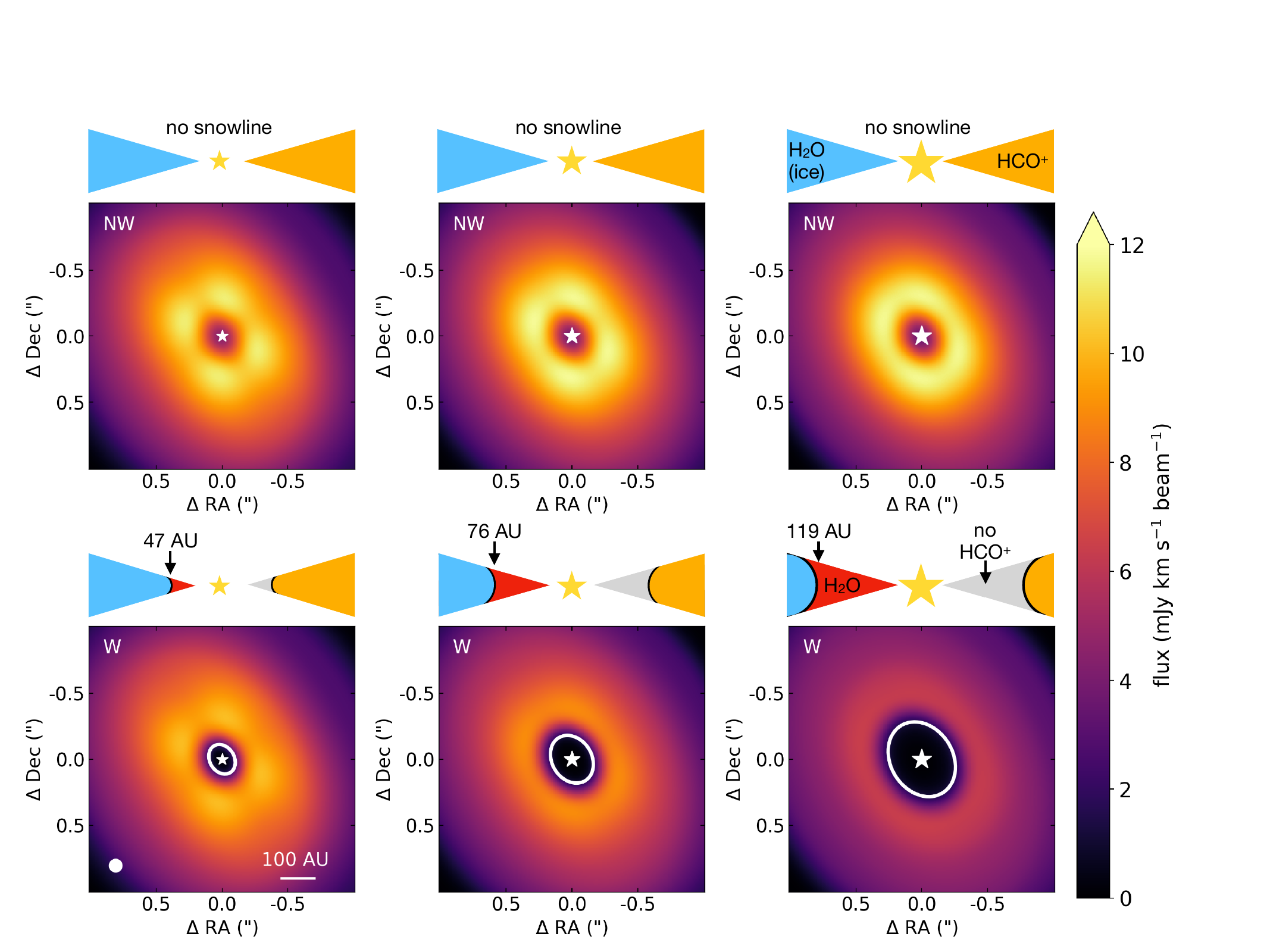}
      \caption{Same as Fig.~\ref{fig:V883Ori_mom0} but then for the $J=2-1$ transition of HCO$^+$. The $0 \farcs 1$ beam and a scale bar are indicated in the bottom left panel.}
         \label{fig:V883Ori_mom0HCO+21}
\end{figure*}

\begin{figure*}
   \centering
   \includegraphics[width=1\linewidth]{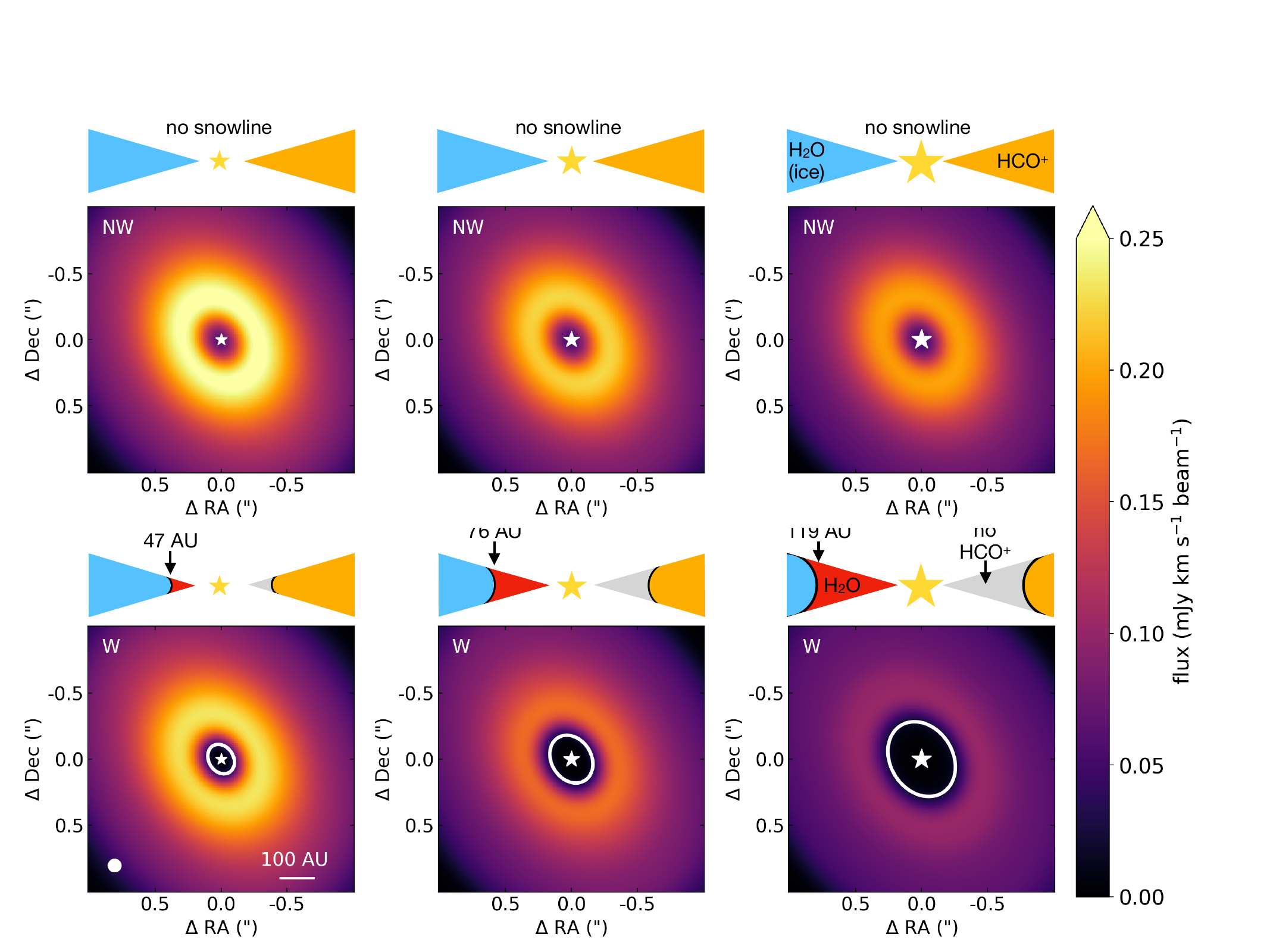}
      \caption{Same as Fig.~\ref{fig:V883Ori_mom0} but then for the $J=2-1$ transition of H$^{13}$CO$^+$. The $0 \farcs 1$ beam and a scale bar are indicated in the bottom left panel.}
         \label{fig:V883Ori_mom0HiCO+21}
\end{figure*}

An overview of the model parameters used for the representative \texttt{DALI} model for V883 Ori is given in Table~\ref{tab:paramsdaliV883}. Predictions for the corresponding emission of the $J=2-1$ transition of HCO$^+$ and H$^{13}$CO$^+$ are shown in Fig.~\ref{fig:V883Ori_mom0HCO+21} and Fig.~\ref{fig:V883Ori_mom0HiCO+21}. 

\begin{table}
\centering
\begin{threeparttable}
	\caption{\texttt{DALI} model parameters for the representative model for V883 Ori. }    
    \begin{tabularx}{\linewidth}{p{0.4\columnwidth}p{0.4\columnwidth}p{0.2\columnwidth}}
    \hline
        Model parameter & Value & Ref. \\ \hline
         \\[-0.7em]
        \textit{Physical structure} &  \\
        $R_{\mathrm{subl}}$ & 1 AU & \\        
        $R_{\mathrm{c}}$ & 75 AU &  \\
		$\Sigma_{\mathrm{c}}$ & 35 g$\ \mathrm{cm^{-2}}$ & \\    
		$M_{\mathrm{disk}}$ & $0.25 \ \mathrm{M_{\odot}}$ & \\   
        $\gamma$ & 1 &  \\
        $h_{\mathrm{c}}$ & 0.1 &  \\
        $\psi$ & 0.25 & \\
 		\\[-0.3em]
        \textit{Dust properties} &  \\        
        $\chi$ & 0.2 & \\
        $f_{\mathrm{ls}}$ & 0.9 & \\
        $\Delta_{\mathrm{gas/dust}}$ & 100 &  \\
		 \\[-0.3em]
        \textit{Stellar spectrum}$^{(1)}$ &  \\
        Type & Outbursting \\		
		$L_{\star + \mathrm{acc}}$ & 2.0(3), 6.0(3), 1.4(4) $\mathrm{L_{\odot}}$ & \\
        $L_{\mathrm{X}}$ & 1.4(30) erg s$^{-1}$ &  \\
        $T_{\mathrm{eff}}$ & $1(4)$ K & \\
        $T_{\mathrm{acc}}$ & $1(4)$ K & \\        
        $T_{\mathrm{X}}$ & 7(7) K & \\          
        $\zeta_{\mathrm{c.r.}}$ & 5(-17) $\mathrm{s^{-1}}$ &\\  
        \\[-0.3em]
        \textit{Stellar properties}$^{(2)}$ &  \\       
        $\dot{M}$ & $5(-5) \ \mathrm{M_{\odot}}$ yr$^{-1}$ & \\      
        $M_{\star}$ & $1.3 \ \mathrm{M_{\odot}}$ & (a) \\
        $R_{\star}$ & $5.1 \ \mathrm{R_{\odot}}$ \\ 
        \\[-0.3em]
        \textit{Observational geometry} &  \\  
        $i$ & 38$\degree$ & (a)\\
        P.A. & 32$\degree$ & (a)\\
        $d$ & 400 pc & (b) \\ \hline
    \end{tabularx}
    \begin{tablenotes}
      \small
      \item \textbf{Notes.} $a(b)$ represents $a\times 10^b$. $^{(1)}$The stellar spectrum is obtained by the sum of the accretion luminosity $L_{\mathrm{acc}}$ and an artificially high stellar luminosity $L_{\star}$ to shift the snowline to 47, 76 and 119~AU. $^{(2)}R_{\star}$ is chosen to obtain an accretion luminosity of $4\times 10^2$~L$_{\odot}$, consistent with \citet{Cieza2016}. References: (a) \citealt{Cieza2016}, (b) \citealt{Kounkel2017}. 
    \end{tablenotes}
    \label{tab:paramsdaliV883}
    \end{threeparttable}
\end{table}

\end{appendix}
\end{document}